\documentclass[aps,preprint,superscriptaddress,nofootinbib,preprintnumbers,prd,eqsecnum]{revtex4-1}

\usepackage{bm, amsmath,amssymb,amsfonts,slashed,graphicx,array,multirow}

\newcommand{\nn}{\nonumber}
\newcommand{\Tr}{\mbox{Tr}\,}

\newcommand{\one}{\bm{1}}
\newcommand{\lmb}{\lambda}
\newcommand{\LambdaH}{\Lambda_{\textrm{H}}}
\newcommand{\LambdaF}{\Lambda_{\textrm{F}}}
\newcommand{\da}{\dagger}
\newcommand{\pda}{{\phantom{\da}}}
\newcommand{\al}{\alpha}
\newcommand{\be}{\beta}
\renewcommand{\hat}{\widehat}
\newcommand{\hal}{{\hat{\alpha}}}
\newcommand{\hbe}{{\hat{\beta}}}
\newcommand{\lmbE}[2]{\langle \lmb^{#1}_{#2} \rangle }

\newcommand\redsout{\bgroup\markoverwith{\textcolor{Magenta}{\rule[0.5ex]{2pt}{1.2pt}}}\ULon}

\usepackage{hyperref}	
\allowdisplaybreaks
\makeatletter
\g@addto@macro\bfseries{\boldmath}
\makeatother

\usepackage[dvipsnames]{xcolor}
\definecolor{red}{rgb}{0.9, 0,0}

\usepackage{hyperref}

\begin{document}
\title{The Flavor-locked Flavorful Two Higgs Doublet Model}

\author{Wolfgang Altmannshofer}
\email{wolfgang.altmannshofer@uc.edu}
\affiliation{Department of Physics, University of Cincinnati, Cincinnati, Ohio 45221, USA}

\author{Stefania Gori}
\email{stefania.gori@uc.edu}
\affiliation{Department of Physics, University of Cincinnati, Cincinnati, Ohio 45221, USA}

\author{Dean J. Robinson}
\email{dean.robinson@uc.edu}
\affiliation{Department of Physics, University of Cincinnati, Cincinnati, Ohio 45221, USA}

\author{Douglas Tuckler}
\email{tuckleds@mail.uc.edu}
\affiliation{Department of Physics, University of Cincinnati, Cincinnati, Ohio 45221, USA}

\begin{abstract}
We propose a new framework to generate the Standard Model (SM) quark flavor hierarchies in the context of two Higgs doublet models (2HDM). The `flavorful' 2HDM couples the SM-like Higgs doublet exclusively to the third quark generation, while the first two generations couple exclusively to an additional source of electroweak symmetry breaking, potentially generating striking collider signatures. We synthesize the flavorful 2HDM with the `flavor-locking' mechanism, that dynamically generates large quark mass hierarchies through a flavor-blind portal to distinct flavon and hierarchon sectors: Dynamical alignment of the flavons allows a unique hierarchon to control the respective quark masses. We further develop the theoretical construction of this mechanism, and show that in the context of a flavorful 2HDM-type setup, it can automatically achieve realistic flavor structures: The CKM matrix is automatically hierarchical with $|V_{cb}|$ and $|V_{ub}|$ generically of the observed size. 
Exotic contributions to meson oscillation observables may also be generated, that may accommodate current data mildly better than the SM itself.
\end{abstract}
\maketitle

\tableofcontents
\clearpage

\section{Introduction}
The LHC collaborations have established with Run I data that the 125 GeV Higgs boson has
Standard Model (SM)-like properties~\cite{Khachatryan:2016vau}.  In particular, the couplings of the Higgs boson to the electroweak gauge bosons have been measured with an uncertainty of $10\%$ at the $1\sigma$ level, 
combining results from ATLAS and CMS~\cite{Khachatryan:2016vau}. The Higgs coupling to $\tau$ leptons has been measured at the $15\%$ level~\cite{Khachatryan:2016vau}, and, assuming no 
significant contribution of new degrees of freedom to the gluon fusion Higgs production cross section, the Higgs coupling to top quarks has been found to be SM-like with approximately $15\%$
uncertainty~\cite{Khachatryan:2016vau}. More recently, analyses of $\sim 36$ fb$^{-1}$ of Run II LHC data have provided evidence for the decay of the Higgs boson into a pair of $b$~quarks 
with a branching fraction consistent with the SM expectation~\cite{Aaboud:2017xsd,Sirunyan:2017elk}. Taken together, these results imply that the main origin of the masses of the weak gauge bosons and third 
generation fermions is the vacuum expectation value (vev) of the 125 GeV SM-like Higgs.
 
However, it is not known whether the vacuum of the SM Higgs field is (solely) responsible for the generation of all the elementary fermion masses. So far, the $h \to \mu\mu$ branching fraction is 
bounded by a factor of $\sim 2.6$ above the SM prediction~\cite{CMS-PAS-HIG-17-019,Aaboud:2017ojs}. With 300~fb$^{-1}$ of data, the SM partial width for this decay mode will be accessible 
at LHC, and it could be measured with a precision of $\sim 8\%$ at the High-Luminosity LHC (HL-LHC)~\cite{CMS:2013xfa,ATL-PHYS-PUB-2013-014,Testa:2017hl}. The $h \to c\bar{c}$ rate is more difficult 
to access at the LHC. At present, the most stringent bound arises from the ATLAS search for $Zh,h\to c\bar c$, exploiting new $c$-tagging techniques, and only probes the branching fraction down to 
$\sim 110$ times the SM expectation~\cite{ATLAS-CONF-2017-078}. Studies of future prospects for the HL-LHC have shown that LHCb may be able to set 
a stronger bound on the $h c\bar c$ coupling, at the level of $\sim 4$ times the SM expectation~\cite{LHCb-CONF-2016-006}. 
The charm coupling may be determined more precisely at future colliders, such as $e^+ e^-$ machines~\cite{Fujii:2017vwa}, as well as proton-electron colliders \cite{Mellado:2015dvt}. 
Finally, because of their tiny values, the SM Higgs couplings to the other light quarks, as well as the electron, are even more challenging to measure and will likely remain out of reach for 
the foreseeable future~\cite{Perez:2015lra,Altmannshofer:2015qra,Bishara:2016jga,Soreq:2016rae,Gao:2016jcm,Bonner:2016sdg,Yu:2016rvv,Carpenter:2016mwd,dEnterria:2017dac,Cohen:2017rsk}. 
Signals that would provide immediate evidence for a beyond SM Higgs sector, such as $h \to \tau\mu$ and $t \to ch$, have branching fractions that are constrained to be less than 
few $\times~ 10^{-3}$ \cite{CMS:2017onh,Aad:2016blu,Aaboud:2017mfd,CMS:2017cck}.
 
At the same time, the origin of the large hierarchies in the SM fermion masses, as well as the hierarchical structure of the CKM quark mixing matrix, is a long-standing open question: the SM flavor puzzle. 
One dynamical approach to this puzzle is to couple the first two generations exclusively to an additional subleading source of electroweak symmetry breaking, in the form of a second Higgs doublet 
or some strong dynamics~\cite{Altmannshofer:2015esa} (see also~\cite{Das:1995df,Blechman:2010cs,Ghosh:2015gpa,Botella:2016krk}). Asserting suitable textures for the quark and 
lepton Yukawa matrices, in order to satisfy flavor constraints, leads to a `flavorful' two Higgs doublet model (F2HDM). The collider signatures of the F2HDM have been explored 
previously~\cite{Altmannshofer:2016zrn}. These include striking signatures for lepton flavor violation, such as $h \to \tau \mu$ or $b \to s\tau \mu$ and large branching ratios for $t \to ch$, 
as well as heavy Higgs or pseudoscalar decays $H/A \to c\bar{c}$, $t\bar{c}$, $\mu\mu$, $\tau\mu$ and charged Higgs decays $H^\pm \to b\bar{c}$, $s\bar{c}$, $\mu \nu$.

A different approach to resolving the SM mass hierarchy puzzle can be achieved with a dynamical alignment mechanism~\cite{Knapen:2015hia} -- we refer to it as `flavor-locking' -- in which
the quark (or lepton) Yukawas are generated by the vacuum of a general flavon potential, that introduces a single flavon field and a single `hierarchon' operator for each quark flavor. 
(A detailed review follows below; see also Refs.~\cite{Cabibbo:1970rza,Michel:1971th,Alonso:2013nca,Crivellin:2016ejn} for related, but somewhat different approaches.) 
In this vacuum,  the up- and down-type sets of flavons are dynamically locked into an aligned, rank-1 configuration in the mass basis, so that each SM quark mass is controlled by a unique flavon. 
Horizontal symmetries between the hierarchon and flavon sectors in turn allow each quark mass to be dynamically set by a unique hierarchon vev. This results in a flavor blind mass generation 
mechanism -- the quarks themselves carry no flavor symmetry beyond the usual $U(3)_{Q,U,D}$ -- so that the quark mass hierarchy can be generated  
independently from the CKM quark mixing hierarchy, by physics that operates at scales generically different to -- i.e. lower than -- the scale of the flavon effective field theory.
In a minimal set up that features only a single SM-type Higgs, however, the CKM mixing matrix is an arbitrary unitary matrix, so that the quark mixing hierarchy itself remains unexplained.

In this work we synthesize these two approaches to the flavor puzzle with the following observation: A dynamical realization of an F2HDM-type flavor structure can be generated by applying 
the flavor-locking mechanism to its Yukawas. Or alternatively: In a flavor-locking scheme for the generation of the quark mass hierarchy, introducing a second Higgs doublet with F2HDM-type 
couplings generically produces quark mixing hierarchies of the desired size. In particular, we show that in such a setup, the $1$--$3$ and $2$--$3$ quark mixings are automatically produced at the 
observed order, without the introduction of tunings. The flavor structure of this theory generically leads to tree-level contributions from heavy Higgs exchange to meson mixing observables, that vanish 
in the heavy Higgs infinite mass limit. However, for heavy Higgs masses at collider-accessible scales, we show these contributions may be consistent with current data, and in some cases may 
accommodate the current data mildly better than the SM.

This paper is structured as follows. In Sec.~\ref{sec:F2HDM} we briefly review the general properties of the F2HDM and its flavor structure. In Sec.~\ref{sec:FL} we develop the flavor-locking mechanism for F2DHM-type theories, including a review of the minimal single Higgs version. We then proceed to explore the generic flavor structure of the flavor-locked F2HDM in Sec.~\ref{sec:FSS}, discussing both the generation of the CKM mixing hierarchies and constraints from meson mixing. We conclude in Sec.~\ref{sec:conclusions}.
Technical details concerning the analysis of the flavon potential are given in Appendices.

\section{Review of the flavorful 2HDM}
\label{sec:F2HDM}
The F2HDM, as introduced in Refs.~\cite{Altmannshofer:2015esa,Altmannshofer:2016zrn}, is a 2HDM in which one Higgs doublet predominantly gives mass to the third generation of quarks and leptons, while the second Higgs doublet is responsible for the masses of the first and second generation of SM fermions, as well as for quark mixing. 
The most general Yukawa Lagrangian of two Higgs doublets with hypercharge $+1/2$ can be written as
\begin{align}
 -\mathcal L_\text{Y} =& \sum_{i,J} \Big[ Y^u_{iJ} (\bar Q_L^i \tilde H_1 U_R^J) + Y^{\prime u}_{iJ} (\bar Q_L^i \tilde H_2 U_R^J) \Big]  + \sum_{i,\hat J} \Big[ Y^d_{i \hat J} (\bar Q_L^i H_1 D_R^{\hat J}) + Y^{\prime d}_{i \hat J} (\bar Q_L^i H_2 D_R^{\hat J}) \Big] \nn \\
 & + \sum_{i,\hat J} \Big[ Y^\ell_{i \hat J} (\bar L_L^i H_1 E_R^{\hat J}) + Y^{\prime \ell}_{i \hat J} (\bar L_L^i H_2 E_R^{\hat J}) \Big]~ + \text{h.c.} ~, \label{LY}
\end{align}
with two Higgs doublets $H_1$ and $H_2$ coupling to the left-handed and right-handed quarks ($Q_L$, $U_R$, $D_R$) and leptons ($L_L$ and $E_R$), and $\tilde H \equiv \epsilon H^*$. The indices $i = 1,2,3$ and $J, \hat J = 1,2,3$ label the three generations of $SU(2)$ doublet and singlet fields, respectively. We focus on quark Yukawas hereafter, but the general results of this discussion apply equally to the lepton Yukawas in Eq.~\eqref{LY}.

The two Higgs doublets decompose in the usual way 
\begin{align}
  H_1 &= \begin{pmatrix}G^+  \sin\beta  - H^+ \cos\beta  \\ \dfrac{1}{\sqrt{2}} ( v \sin\beta + h \cos\alpha + H \sin\alpha + i G^0 \sin\beta - i A \cos\beta) \end{pmatrix}  \, , \\
  H_2 &= \begin{pmatrix} G^+ \cos\beta + H^+ \sin\beta \\ \dfrac{1}{\sqrt{2}} ( v \cos\beta - h \sin\alpha + H \cos\alpha + i G^0 \cos\beta + i A \sin\beta) \end{pmatrix} \, ,
\end{align}
where $v = 246$~GeV is the vacuum expectation value of the SM Higgs, $G^0$ and $G^\pm$ are the Goldstone bosons that provide the longitudinal components for the $Z$ and $W^\pm$ bosons, $h$ and $H$ are physical scalar Higgs bosons, $A$ is a physical pseudoscalar Higgs boson, and $H^\pm$ are physical charged Higgs bosons. The angle $\alpha$ parametrizes diagonalization of the scalar Higgs mass matrix and $\tan\beta$ is the ratio of the vacuum expectation values of $H_1$ and $H_2$.
The scalar $h$ is identified with the 125~GeV Higgs boson. The overall mass scale of the `heavy' Higgs bosons $H,A,H^\pm$ is a free parameter. The mass splitting among them is at most of order $O(v^2/m_{H,A,H^\pm})$. 

In Refs.~\cite{Altmannshofer:2015esa,Altmannshofer:2016zrn} the following textures of the two sets of Yukawa couplings $Y$ and $Y^\prime$ were chosen,
\begin{subequations}
\renewcommand*{\arraystretch}{.75}
\label{eqn:Ys}
\begin{align} \label{eq:Yu}
 Y^u &\sim \frac{\sqrt{2}}{v \sin\beta} \begin{pmatrix} 0 &  &  \\  & 0 &  \\  &  & m_t \end{pmatrix} \,, & Y^{\prime u} & \sim \frac{\sqrt{2}}{v \cos\beta} \begin{pmatrix} m_u & m_u & m_u \\ m_u & m_c & m_c \\ m_u & m_c & m_c \end{pmatrix} \,, \\ \label{eq:Yd}
 Y^d &\sim \frac{\sqrt{2}}{v \sin\beta} \begin{pmatrix} 0 &  &  \\  & 0 &  \\  &  & m_b \end{pmatrix} \,, & Y^{\prime d} & \sim \frac{\sqrt{2}}{v \cos\beta} \begin{pmatrix} m_d & \lambda m_s & \lambda^3 m_b \\ m_d & m_s & \lambda^2 m_b \\ m_d & m_s & m_s \end{pmatrix} \,, 
\end{align}
\end{subequations}
where each entry in the $Y^{\prime u},Y^{\prime d}$ Yukawas is multiplied by a generic $\mathcal O(1)$ coefficient. This structure naturally produces the observed quark
masses as well as CKM mixing angles. In this work, we will focus on the \emph{dynamical} generation of Yukawas of a similar form, with the schematic structure
\begin{subequations}
\label{eqn:YsFL}
\renewcommand*{\arraystretch}{.75}
\begin{align}
	Y^u & \sim \frac{\sqrt{2}}{v \sin\beta} \begin{pmatrix} 0  && \\ & 0 & \\ && m_t \end{pmatrix}\,,  &  Y^{\prime u} & \sim  \frac{\sqrt{2}}{v \cos\beta} U_u \begin{pmatrix} m_u   && \\ & m_c & \\ && 0 \end{pmatrix} V_u^\da \,,\\
	Y^d & \sim \frac{\sqrt{2}}{v \sin\beta} \begin{pmatrix} 0  && \\ & 0 & \\ && m_b \end{pmatrix}\,,  &  Y^{\prime d} & \sim  \frac{\sqrt{2}}{v \cos\beta} U_d \begin{pmatrix} m_d   && \\ & m_s & \\ && 0 \end{pmatrix} V_d^\da\,,
\end{align}
\end{subequations}
in which $U_{u,d}$ and $V_{u,d}$ are unitary matrices. These Yukawas will similarly produce the observed quark mass hierarchies and CKM mixing (see Sec.~\ref{sec:FSS} below), and the collider phenomenology of both Yukawa structures is expected to manifest in the same set of signatures. 

The F2HDM setup exhibits a very distinct phenomenology, that differs significantly from 2HDMs with natural flavor conservation, flavor alignment, or minimal flavor violation~\cite{Glashow:1976nt,Pich:2009sp,Altmannshofer:2012ar,Gori:2017qwg,DAmbrosio:2002vsn}.
The couplings of the 125~GeV Higgs are modified in a flavor non-universal way. In particular, in regions of parameter space where the couplings of $h$ to the third generation are approximately SM like, the couplings to the first and second generation can still deviate from SM expectations by an $\mathcal{O}(1)$ factor. 
Also, the heavy Higgs bosons $H$, $A$, and $H^\pm$ couple to the SM fermions in a characteristic flavor non-universal way. Their couplings to the third generation are suppressed by $\tan\beta$, while the couplings to first and second generation are enhanced by $\tan\beta$. Therefore, the decays of $H$, $A$, and $H^\pm$ to the third generation -- $t$, $b$ quarks and the $\tau$ lepton -- are not necessarily dominant. 
For large and moderate $\tan\beta$ we expect sizable branching ratios involving, for example, charm quarks and muons. 
Similarly, novel non-standard production modes of the heavy Higgs bosons involving second generation quarks can be relevant and sometimes even dominant~\cite{Altmannshofer:2016zrn}.

One important aspect of the Yukawa structures in Eqs.~\eqref{eqn:Ys} and Eqs.~\eqref{eqn:YsFL} is that they imply tree-level flavor changing neutral Higgs couplings. The flavor-violating couplings of the 125~GeV Higgs vanish in the decoupling/alignment limit, i.e. for $\cos(\beta - \alpha) = 0$. However, flavor-violating couplings of the heavy Higgs bosons persist in this limit and they are proportional to $\tan\beta$. Therefore, for large $\tan\beta$ and heavy Higgs boson masses below the TeV scale, flavor violating processes, such as meson mixing, constrain the F2HDM parameter space.
Note that the rank-1 nature of the third generation Yukawas, $Y$, preserves a $U(2)^5$ flavor symmetry acting on the first and second generation of fermions. This symmetry is only broken by the $Y^\prime$ Yukawa
couplings of the second doublet, so that flavor changing transitions from the second to the first generation are
protected. Therefore, the constraints from kaon and $D$-meson oscillation will be less stringent than one might naively expect. We will discuss meson oscillation constraints in detail in Sec.~\ref{sec:FSS}.

\section{Flavor-Locking with one and two Higgs bosons}
\label{sec:FL}
While the distinct phenomenology of the F2HDM alone motivates detailed studies, a mechanism that realizes the flavor structure in Eqs.~\eqref{eqn:Ys} or~\eqref{eqn:YsFL} has not been explicitly constructed so far. We now discuss how the flavor structure~\eqref{eqn:YsFL} can be dynamically generated by the flavor-locking mechanism, and, conversely, how a F2HDM-type theory permits the flavor-locking mechanism to generate realistic flavor phenomenology. We first review the minimal single Higgs doublet version of the flavor-locking mechanism, followed by the generalization to a theory with two Higgs doublets in Sec.~\ref{ref:2H_locking}. As we will discuss, while in the presence of only one SM-like Higgs doublet, the predicted quark mixing angles are generically of $\mathcal O(1)$, introducing a second Higgs doublet leads to a theory with suppressed $|V_{cb}|$ and $|V_{ub}|$.

\subsection{Yukawa portal}
The underlying premise of the flavor-locking mechanism~\cite{Knapen:2015hia} is that the Yukawas arise from a three-way portal between the SM fields (the quarks $Q_L, U_R, D_R$ and the Higgs $H$),
a set of `flavon' fields, $\lmb$, and a set of `hierarchon' operators, $s$:
\begin{equation}
	\label{eqn:3WP}
	-\mathcal L_\text{Y} \supset \bar{Q}_L^i \frac{{\lmb_\al}_{iJ}}{\LambdaF} \frac{s_\al}{\LambdaH} \tilde H U_R^J + \bar{Q}_L^i \frac{{\lmb_\hal}_{i\hat J}}{\LambdaF} \frac{s_\hal}{\LambdaH} H D_R^{\hat J}\,.
\end{equation}
The $\lmb$'s are bifundamentals of the appropriate $U(3)_Q\times U(3)_{U,D}$ flavor groups for up and down quarks, respectively. The subscripts\footnote{We always distinguish down-type indices from up-type indices with a hat, and similarly for down-type versus up-type flavon couplings and operators.}, $\al = u, c, t$ and $\hal = d, s, b$, denote an arbitrary transformation property under a symmetry or set of symmetries, $\mathcal{G}$ and $\hat{\mathcal{G}}$, that enforces the structure of Eq.~\eqref{eqn:3WP}. In the original flavor-locking study~\cite{Knapen:2015hia}, $\mathcal{G}\times\hat{\mathcal{G}}$ was chosen to be a set of discrete $\mathbb{Z}^{p_q}_q$ or $U(1)_{q}$ `quark flavor number' symmetries, for $q = d,s,b,u,c,t$. Here, we similarly choose each flavon $\lmb_{\al}$ ($\lmb_{\hat\al}$) to be charged under a gauged $U(1)_{\al}$ ($U(1)_{\hat\al}$), but assert a $S_3$ permutation symmetry among the up (down) flavons and the corresponding $U(1)_{\al}$ ($U(1)_{\hal}$) gauge bosons, fixing the gauge 
couplings $g _\al = g$ ($g_{\hal} = \hat g$). 
Compared to the analysis of Ref.~\cite{Knapen:2015hia} the permutation symmetry produces a convenient, higher symmetry for the flavon potential, such that configurations with the structure of Eqs.~\eqref{eqn:YsFL} can be shown to be at its global minimum, as we will discuss in the next subsection.
Note that the SM fields are not charged under the $\mathcal{G}\times\hat{\mathcal{G}}$ symmetry.

The hierarchons $s$ should be thought of as some set of scalar operators that eventually obtain hierarchical vevs, that break the $S_3$ symmetries in the up and down sectors. This hierarchy will be responsible for the quark mass hierarchy, independently from any flavor structure. It should be emphasized that the operators $s_\al$ and $s_\hal$ do not carry the quark $U(3)_Q\times U(3)_{U,D}$ flavor symmetries, i.e., they do not carry flavor indices $i,J,\hat J$. Moreover, the hierarchon scale $\LambdaH$ need not be the same as the flavon scale $\LambdaF$, and can generically be much lower. (This could permit, in principle, collider-accessible hierarchon phenomenology, depending on the UV completion of the hierarchon sector, though we shall not consider such possibilities in this work.)

In the remainder of this section, we present the general flavor structures that this type of portal dynamically produces. Details of this analysis, including the identification of global or local minima of the flavon potential, and the algebraic structure of the associated vacua, are presented in Appendix~\ref{app:GFPGM}.
The spontaneous breaking of continuous symmetries by the flavon vacuum can result in a large number of Goldstone bosons.
We assume that mechanisms are at work that remove the Goldstone bosons from the IR. 

\subsection{General flavon potential and vacuum}
To generalize beyond the three flavors of the SM, we contemplate a theory of $N$ flavors of up and down type quarks each, $Q_L^i, U_R^J, D_R^{\hat J}$ with $i,J,\hat J = 1,\dots, N$, charged under the symmetry $U(N)_Q \times U(N)_U \times U(N)_D$. 
We introduce $n \leq N$ pairs of flavons $\lambda_\alpha$, $\lambda_{\hat \alpha}$, with $\alpha, \hat\alpha = 1,\dots,n$, that generate Yukawa couplings to the quarks as in Eq.~\eqref{eqn:3WP}.
The flavons for this theory then transform as
\begin{equation}
	\lmb_\al \sim \bm{N}\otimes \bm{\bar{N}} \otimes \one\,,\qquad \lmb_\hal \sim \bm{N}\otimes \one \otimes \bm{\bar{N}}\,.
\end{equation}
We suppress hereafter the $U(N)_Q\times U(N)_{U,D}$ indices, keeping in mind that 
matrix products only take the form $\lmb_\al^\pda\lmb_\be^\da$ or $\lmb_\be^\da\lmb_\al^\pda$, and correspondingly in the down sector. 
Up-down matrix products can only take the form $\lmb_\al^\da\lmb_\hal^\pda$ or $\lmb_\hal^\da\lmb_\al^\pda$, but not $\lmb_\al^\pda\lmb_\hal^\da$ nor $\lmb_\hal^\pda\lmb_\al^\da$.

The most general, renormalizable and CP conserving potential for the flavons can then be written in the form
\begin{equation}
	\label{eqn:FP}
	V_{\text{fl}} = \sum_\al V^{\al}_{\text{1f}} + \sum_{\al < \be} V^{\al\be}_{\text{2f}} + \sum_\hal V^{\hal}_{\text{1f}}  + \sum_{\hal < \hbe}V^{\hal\hbe}_{\text{2f}}  + \sum_{\al,\,\hal}V^{\al\hal}_{\text{mix}}\,.
\end{equation}
Here, the single and pairwise field potentials are
\begin{align}\label{eq:SP}
	V^{\al}_{\text{1f}} 
		& = \mu_1\Big| \Tr\big(\lmb_\al^\pda \lmb_\al^\da\big) - r^2  \Big|^2 + \mu_2\Big[\big| \Tr\big(\lmb_\al^\pda \lmb_\al^\da\big)\big|^2 - \Tr\big(\lmb^\pda_\al \lmb^\da_\al\lmb^\pda_\al \lmb^\da_\al\big) \Big]\,,\\
	V^{\al\be}_{\text{2f}} 
		& = \mu_3\Big| \Tr\big(\lmb_\al^\pda \lmb_\al^\da\big) - \Tr\big(\lmb_\be^\pda \lmb_\be^\da\big)\Big|^2 + \mu_4 \Big|\Tr\big(\lmb_\al^\pda \lmb_\be^\da\big) \Big|^2\nn \\
		& \qquad + \mu_{6,1}\Tr\big(\lmb^\da_\al \lmb^\pda_\al\lmb^\da_\be \lmb^\pda_\be\big)   +  \mu_{6,2}\Tr\big(\lmb^\pda_\al \lmb^\da_\al\lmb^\pda_\be \lmb^\da_\be\big)\,, \label{eqn:SPFP}
\end{align}
and similarly for $V^{\hal}_{\text{1f}}$ and $V^{\hal\hbe}_{\text{2f}}$, hatting all coefficients (the labeling and notation follows the choices of Ref.~\cite{Knapen:2015hia}). 
Note that the pairwise potentials respect the $U(1)_{\al}$ and $U(1)_{\hat\al}$ symmetries.
The mixed potential is
\begin{align}
	V^{\al\hal}_{\text{mix}} 
		& = \nu_1r^2 \hat r^2 \Big| \Tr\big(\lmb_\al^\pda \lmb_\al^\da\big)/r^2 - \Tr\big(\lmb_\hal^\pda \lmb_\hal^\da\big)/\hat r^2\Big|^2 \nn \\
		& \qquad - \nu_2\bigg[\Tr\big(\lmb^\pda_\al \lmb^\da_\al\lmb^\pda_\hal \lmb^\da_\hal\big) -  \frac{1}{n} \Tr\big(\lmb_\al^\pda \lmb_\al^\da\big)\Tr\big(\lmb_\hal^\pda \lmb_\hal^\da\big) \bigg]\,.\label{eqn:VM}
\end{align}
The $S_n$ symmetry ensures that all potential coefficients are the same for all fields $\alpha,\hat\alpha,\beta,\hat\beta$ singly and pairwise. All $\mu_i$ and $\nu_i$ coefficients, as well as $r$ and $\hat r$, are real and are chosen to be positive.

A detailed analysis of the global minimum of this potential is provided in Appendix~\ref{app:GFPGM}. One finds that, provided 
\begin{equation}
	\label{eqn:CGMA}
	\mu_{6,2} \ge \nu_2 \hat r^2/ r^2\,, \qquad \hat\mu_{6,2} \ge \nu_2 r^2/\hat r^2\,, \qquad \text{and} \qquad \nu_1 \ge \nu_2/(2 n)\,,
\end{equation}
the potential has a global minimum if and only if the flavons have the vacuum configuration
\begin{subequations}
\renewcommand*{\arraystretch}{.7}
\label{eqn:FLC}
\begin{align}
	\langle \lmb_1 \rangle & = U \begin{pmatrix} r  && \\ & 0 & \\[-5pt]  && \ddots \end{pmatrix} V^\da\,,&  
	\langle \lmb_2 \rangle & =  U\begin{pmatrix} 0  && \\ & r & \\[-5pt] && \ddots \end{pmatrix} V^\da\,, & \ldots \\
	\langle \lmb_{\hat{1}} \rangle & = \hat{U} \begin{pmatrix} \hat{r}  && \\ & 0 & \\[-5pt]  && \ddots \end{pmatrix} \hat{V}^\da\,,&  
	\langle \lmb_{\hat{2}} \rangle & =  \hat{U} \begin{pmatrix} 0  && \\ & \hat{r} & \\[-5pt]  && \ddots \end{pmatrix} \hat{V}^\da\,, & \ldots  \label{eqn:FLVL}
\end{align}
\end{subequations}
with $U$, $V$, $\hat{U}$, $\hat{V}$ unitary matrices -- crucially, the matrices $U$, $V$ ($\hat{U}$, $\hat{V}$) are the same for all $\lmb_\al$ ($\lmb_\hal$) -- and the CKM mixing matrix has the form
\begin{equation}
	\label{eqn:MGCKMA}
	\mathcal{V}_{\text{ckm}} = U^\da\hat{U} = \begin{pmatrix} \mathcal{V}_n & 0 \\ 0 &  \mathcal{V}_{N-n} \end{pmatrix}\,,
\end{equation}
with $\mathcal{V}_k$ a $k\times k$ unitary matrix. These $n$ or $N-n$ block CKM rotations are flat directions of the global minimum, and therefore $\mathcal{V}_n$ and $\mathcal{V}_{N-n}$ may be any arbitrary unitary submatrices with generically $\mathcal O(1)$ entries. We refer to the configuration in Eqs.~\eqref{eqn:FLC} and \eqref{eqn:MGCKMA} as being `flavor-locked'.

\subsection{Flavor-locked Yukawas}
Flavor locking ensures that the Yukawa portal in~\eqref{eqn:3WP} becomes, in the $n=N=3$ case
\begin{equation}
	\label{eqn:CKMSH}
	\bar{Q}_L \frac{r}{\LambdaF} \begin{pmatrix} s_u/\LambdaH && \\ & s_c/\LambdaH & \\ && s_t/\LambdaH \end{pmatrix} \tilde H U_R + \bar{Q}_L \frac{\hat{r}}{\LambdaF} \mathcal{V}_{\text{ckm}} \begin{pmatrix} s_d/\LambdaH && \\ & s_s/\LambdaH & \\ && s_b/\LambdaH \end{pmatrix} H D_R\,,
\end{equation}
under a suitable unitary redefinition of the $Q_L,U_R$ and $D_R$ fields. From these expressions, taking the natural choice $r, \hat r \sim \LambdaF$, it is clear that it is the physics of the hierarchon vev's, $\langle s_\al \rangle$, that generates the quark mass hierarchies, i.e. $\langle s_\al \rangle/\LambdaH \sim y_\al$, the quark Yukawa for flavor $\al$. This physics may operate at scales vastly different to the flavor breaking scale, $\LambdaF$. In Eq.~\eqref{eqn:CKMSH} the CKM matrix $\mathcal{V}_{\text{ckm}}$ is an arbitrary $3\times 3$ unitary matrix.

One might wonder if additional terms in the flavon potential of (\ref{eqn:FP}) can destabilize the vacuum identified above. 
In particular, flavon-hierarchon couplings of the form $\Tr[\lmb^\da_\al \lmb^\pda_\al] s^\da_\al s^\pda_\al$ ($\Tr[\lmb^\da_\al \lmb^\pda_\be] s^\da_\al s^\pda_\be$) may be present, which can produce (mixed) mass terms that disrupt the $V_{\text{mix}}$ ($V_{\text{2f}}$) vacuum once the hierarchons, $s_\al$, obtain vev's. Mixed mass terms may disrupt the alignment between the different $\lmbE{}{\al}$, while additional mass terms induce splittings in the radial mode masses, so that the block CKM rotations are no longer flat directions of the vacuum.

In the UV theory, the operator product of two hierarchons with two flavons may, however, be vanishingly small, e.g. if the hierarchons are composite operators in different sectors. Nonetheless, such terms are necessarily generated radiatively by the Yukawa portal~\eqref{eqn:3WP}. One may construct UV completions in which this occurs first at the two-loop level, with the (mixed) mass contributions being log-divergent. For example, let us consider a theory containing a flavored fermion $\chi_{\al i}$ and a scalar $\Phi_\al$, with interactions
\begin{equation}
	\label{eqn:UVC}
	\lmb_{\al i J} \bar\chi_{\al i} U_R^J + \Phi^\pda_\al \bar{Q}_{L}^i \chi_{\al i}   + \mu \Phi^\da_\al s_\al \tilde H\,,
\end{equation}
with $m_\chi \sim \LambdaF$ and $\mu \sim m_\Phi \sim \LambdaH$. This produces the Yukawa portal~\eqref{eqn:3WP} via
\begin{equation}\label{eq:diagramUV}
	\parbox{0.7\linewidth}{
	\begin{center}
	\includegraphics[width = 5cm]{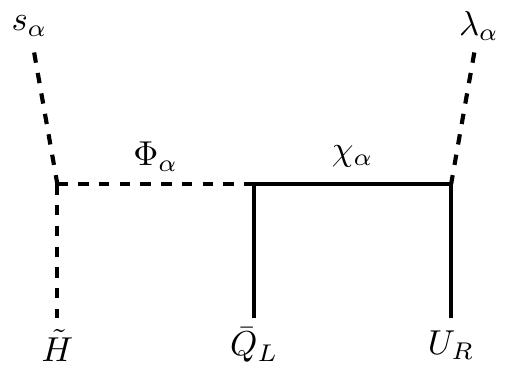}
	\end{center}
	}
\end{equation}
As $\langle s_\al \rangle/\LambdaH \sim y_\al$, the quark Yukawa for flavor $\al$, the corresponding (mixed) mass term for the flavons is generated at two-loops by mirroring the diagram in \eqref{eq:diagramUV}. One finds
\begin{equation}
	\delta m^2_{\al\be} \sim \frac{\LambdaH^2}{\LambdaF^2} \frac{y_\al y_\be}{(16 \pi^2)^2} \log(\LambdaH/\LambdaF)\, r^2\,,
\end{equation}
once again taking the natural choice $r \sim \LambdaF$. A suitable hierarchy between $\LambdaH$ and $\LambdaF$, combined with the two-loop suppression, renders these terms arbitrarily small. Hence one may safely neglect these terms.

\subsection{Two-Higgs flavor-locking} 
\label{ref:2H_locking}
Motivated by the flavorful 2HDM, now we turn to consider a Yukawa potential with two Higgs fields: One that couples to the third generation, and one to the first two generations. That is,
\begin{equation}
	\label{eqn:2H3WP}
	\bar{Q}_L \bigg[\frac{{\lmb_t}}{\LambdaF} \frac{s_t}{\LambdaH}\tilde H_1 +  \frac{{\lmb_{c,u}}}{\LambdaF} \frac{s_{c,u}}{\LambdaH} \tilde H_2 \bigg] U_R + \bar{Q}_L \bigg[\frac{{\lmb_b}}{\LambdaF} \frac{s_b}{\LambdaH} H_1 + \frac{{\lmb_{s,d}}}{\LambdaF} \frac{s_{s,d}}{\LambdaH} H_2 \bigg] D_R\,,
\end{equation}
in which we have suppressed the quark flavor indices. With reference to the UV completion~\eqref{eqn:UVC}, one can imagine that this generational structure comes about as a consequence 
of $\lmb_t$, $s_t$, and $H_1$ belonging to a different UV sector (or brane) than $\lmb_{c,u}$, $s_{c,u}$, and $H_2$, so that terms of the form $\lmb_t s_t \tilde H_2$ or $\lmb_{c,u}s_{c,u} \tilde H_1$ 
are heavily suppressed in the effective field theory. Similarly, one can also generate this structure via adding an additional symmetry to $s_{c,u}$, $s_{s,d}$ and $H_2$ such that $s_{c,u} \tilde H_2$ 
and $s_{d,s} H_2$ are singlets. Such terms (symmetries) will, ultimately, be generated (softly broken) via the $\mu^2 H_1 H_2^\da$ term in the Higgs potential, which is necessary to avoid 
a massless Goldstone boson.

The generational structure implies that cross-terms between the third and first two generations in the flavon potential~\eqref{eqn:FP} now vanish, and that the $S_3$ flavon-hierarchon symmetry 
has been replaced with a $\mathbb{Z}_2$ for just the two light generations. That is, the coefficients of the heavy and light flavon potentials are no longer related, and the heavy-light potentials 
$V^{t\al}_{\text{2f}}$, $V^{b\hal}_{\text{2f}}$, $V^{t \hal}_{\text{mix}}$, $V^{b\al}_{\text{mix}}$ vanish, for $\al = c,u$ and $\hal = s,d$ (or they obtain their own, independent, and suppressed coefficients, 
identical for $\al = c,u$ and $\hal = s,d$). One then also expects the rotation matrices entering in the vacuum configuration of the flavons of the first two generations to be different from those of the third, breaking 
the heavy-light alignment conditions.

Put a different way, we may write the full potential in the form
\begin{equation}
	\label{eqn:THFP}
	V_{\text{fl}} = V_{\text{fl}, \text{h}} + V_{\text{fl},\text{l}}
\end{equation}
in which the `$\text{h}$' and `$\text{l}$' pieces of the potential each have the form of the full potential~\eqref{eqn:FP}, but for one heavy and two light generations, respectively. (With reference to the UV completion~\eqref{eqn:UVC}, terms for a heavy-light mixing potential are generated radiatively by the $\mu^2 H_1 H_2^\da$ portal combined with the Yukawas~\eqref{eqn:2H3WP} only at the five-loop level, along with a $\mu^4/\LambdaF^4$ factor.)
The potentials $V_{\text{fl}, \text{h}}$ and $V_{\text{fl},\text{l}}$ each have a $N=3$ flavor-locked vacuum, with generation number $n=1$ and $n=2$, respectively. 
Provided the conditions~\eqref{eqn:CGMA} are satisfied for each potential, this leads to the vacuum structure
\renewcommand*{\arraystretch}{.6}
\begin{align}
	\langle \lmb_t \rangle & = U^\pda_t \begin{pmatrix} 0  && \\ & 0 & \\ && r \end{pmatrix} V_t^\da\,,&  
	\langle \lmb_{c} \rangle & =  U\begin{pmatrix} 0  && \\ & r & \\ && 0 \end{pmatrix} V^\da\,,& 
	\langle \lmb_{u} \rangle & =  U \begin{pmatrix} r  && \\ & 0 & \\ && 0 \end{pmatrix} V^\da\,,\nn \\
	\langle \lmb_b \rangle & = \hat{U}^\pda_b \begin{pmatrix} 0  && \\ & 0 & \\ && \hat{r} \end{pmatrix} \hat{V}_b^\da\,,&  
	\langle \lmb_{s} \rangle & =  \hat{U} \begin{pmatrix} 0  && \\ & \hat{r} & \\ && 0 \end{pmatrix} \hat{V}^\da\,,& 
	\langle \lmb_{d} \rangle & =  \hat{U} \begin{pmatrix} \hat{r}  && \\ & 0 & \\ && 0 \end{pmatrix} \hat{V}^\da\,.\label{eqn:21FV}
\end{align}
We call this a `$1+2$' flavor-locked vacuum. Note that the rotation matrices for the third generation quarks ($U_t, V_t, \hat U_b, \hat V_b$) differ in 
general from the corresponding rotations for the first and second generation quarks.

For the $1+2$ flavor-locked structure~\eqref{eqn:21FV}, the CKM structure of the global minimum in Eq.~\eqref{eqn:MGCKMA} enforces  $U^\da\hat{U}$ and $U_t^\da\hat{U}_b$ to each be $2\oplus 1$ block unitary, i.e.
\begin{equation}
	\label{eqn:UUH}
	U^\da\hat{U} = \begin{pmatrix} \mathcal{V}_2 & \\  & 1 \end{pmatrix}\,, \qquad U^\da_t \hat{U}^\pda_b = \begin{pmatrix} \mathcal{W}_{2} &  \\  & 1 \end{pmatrix}\,,
\end{equation}
where $\mathcal{V}_2$ and $\mathcal{W}_2$ are $2 \times 2$ unitary matrices (see App.~\ref{app:AFG}).  The $2 \oplus 1$ block unitarity permits one to rotate away the $tb$ unitary matrices, so that the Yukawa potential~\eqref{eqn:2H3WP} attains the form 
\begin{multline}
	\bar{Q}_L \frac{r}{\LambdaF} \Bigg[ \begin{pmatrix} 0  && \\ & 0 & \\ && z_t \end{pmatrix} \tilde H_1 +  U \begin{pmatrix} z_u   && \\ & z_c & \\ && 0 \end{pmatrix} V^\da \tilde H_2 \Bigg] U_R \\
	+ \bar{Q}_L \frac{\hat{r}}{\LambdaF} \Bigg[ \begin{pmatrix} 0  && \\ & 0 & \\ && z_b \end{pmatrix} H_1 + U  \begin{pmatrix} \mathcal{V}_2 & \\  & 1 \end{pmatrix} \begin{pmatrix} z_d   && \\ & z_s & \\ && 0 \end{pmatrix} \hat{V}^\da H_2 \Bigg] D_R\,, \label{eqn:YPR}
\end{multline}
with $z_{\alpha} = \langle s_\alpha \rangle/\LambdaH$ and $z_{\hal} = \langle s_\hal \rangle/\LambdaH$.
The unitary matrices $U$, $V$ and $\hat{V}$ have been redefined to absorb the other unitary matrices, such that Eq.~\eqref{eqn:UUH} is still satisfied, and we have written $\hat U = U \text{diag}\{\mathcal{V}_2 ,1\}$ accordingly. Matching the structure of Eq.~\eqref{eqn:YsFL}, Eq.~\eqref{eqn:YPR} is the key result of this section: The dynamical generation of hierarchical aligned third generation Yukawas, and hierarchical aligned first two generation Yukawas. An additional feature, not present in Eq.~\eqref{eqn:YsFL}, is that the up- and down-type light Yukawas are aligned up to an overall mixing angle on the left. The mixing angle is a flat direction of the flavon potential and therefore generically of $O(1)$.

\section{Flavor violation and phenomenology}
\label{sec:FSS}
We now turn to examine the phenomenology of flavor-violating processes generated by the Yukawa structure in Eq.~\eqref{eqn:YPR}. If one treats the SM as a UV complete theory, then the quark sector alone naively features multiple tunings towards the infinitesimal: five for the masses of all quarks except the top, and two for the small size of $|V_{cb}|$ and $|V_{ub}|$. In the minimal or F2HDM-type flavor-locking scenarios, the quark mass hierarchies no longer require such tunings, as they can be generated dynamically by $\langle s_\alpha \rangle$. We show below that the structure of Eq.~\eqref{eqn:YPR} also characteristically produces $1$--$3$ and $2$--$3$ quark generation mixing comparable to the observed size of $|V_{cb}|$ and $|V_{ub}|$, without requiring ad hoc suppression of the underlying parameters. In this sense of counting tunings, the flavor-locked F2HDM is a more natural 
theory of flavor than the SM. Additionally, for the flavor structure~\eqref{eqn:YPR}, the heavy Higgs bosons may remain light enough to be accessible to colliders, i.e. with a mass of a few hundred GeV, while not introducing unacceptably large tree-level contributions to meson mixing observables. In some regions of parameter space, these additional contributions better accommodate the current data than the SM. We explore the nature of such contributions below.

\subsection{Physical parameters}
Starting from the general structure of Eq.~\eqref{eqn:YPR}, which has already selected the direction of the $H_1$-generated component of the third generation, the $Q$, $U$ and $D$ quarks have a maximal $U(2)^3\times U(1)$ flavor symmetry, which breaks to baryon number. This corresponds to $3$ real and $9$ imaginary broken generators. The up-type Yukawa in Eq.~\eqref{eqn:YPR} has a total of $3 + 3 + 3 = 9$ real parameters ($z_{t,u,c}$, and the $SO(3)$ rotations of $U$ and $V$) and $6 + 6 - 2 - 2 = 8$ imaginary parameters (the phases of $U$ and $V$, less the phases commuted or annihilated by the rank-2 diagonal matrix).
The down-type Yukawa, excluding parameters already contained in $U$, has $3 + 1 + 3  = 7$ real parameters ($z_{b,d,s}$, and the $SO(2)$ and $SO(3)$ rotations of $\mathcal V_2$ and $\hat V$, respectively) and $3 + 6 - 2 -1 = 6$ imaginary parameters (the phases of $\mathcal V_2$ and $\hat V$, less the phases commuted or annihilated by the rank-2 diagonal matrix). This counting implies that the total number of physical parameters is $9+8+7+6-12=18$, corresponding to $6$ masses, $7$ angles and $5$ phases. 

To see this explicitly, we write a general $3\times 3$ unitary matrix in the canonical form
\begin{equation}
	U = \begin{pmatrix} e^{i\phi_1} && \\ & e^{i\phi_2} & \\ & & 1 \end{pmatrix} R_U(\theta_{12}) R_U(\theta_{13}, \phi) R_U(\theta_{23})  \begin{pmatrix} e^{i\phi_4} && \\ & e^{i\phi_5} & \\ & & e^{i\phi_6}\end{pmatrix} \,,
\end{equation}
with $R_U$ rotation matrices in the $3\times 3$ flavor space, and $\theta_{12},\theta_{13},\theta_{23}$ and $\phi,\phi_{1,2,4,5,6}$ generic angles and phases, respectively. Here the indices of the angles label the $2 \times 2$ rotations. After redefining several phases, we obtain the parametrization
\begin{multline}
	\bar{Q}_L \frac{r}{\LambdaF} \Bigg[ \begin{pmatrix} 0  && \\ & 0 & \\ && z_t \end{pmatrix} \tilde H_1 +   R_U(\theta_{13},0) R_U(\theta_{23}) \begin{pmatrix} z_u e^{i\psi_u}   && \\ & z_c e^{i\psi_c} & \\ && 0 \end{pmatrix} R^\da_V(\vartheta_{23}) R^\da_V(\vartheta_{13},0) \tilde H_2 \Bigg] U_R \\
	+ \bar{Q}_L \frac{\hat{r}}{\LambdaF} \Bigg[ \begin{pmatrix} 0  && \\ & 0 & \\ && z_b \end{pmatrix} H_1 + R_U(\theta_{13},0) R_U(\theta_{23}) \begin{pmatrix} e^{i\psi_m}\!\!\!\! && \\ & 1 & \\ & & 0\end{pmatrix} \begin{pmatrix} R(\theta) & \\  & 1 \end{pmatrix} \\
	\times \begin{pmatrix} z_d e^{i\psi_d}   && \\ & z_s e^{i\psi_s} & \\ && 0 \end{pmatrix} R^\da_{\hat V}(\hat\vartheta_{23}) R^\da_{\hat V}(\hat \vartheta_{13},0) H_2 \Bigg] D_R\,.
	\label{eqn:PPMM}
\end{multline}
There is a flavor basis in which the above parametrization reproduces the F2HDM textures shown in~(\ref{eqn:Ys}), with coefficients that depend on the several angles $\theta,\vartheta,\hat\vartheta$.
In Appendix~\ref{app:textures} we show explicitly how to rotate into this flavor basis.

\subsection{CKM phenomenology}
The unitary $\mathcal{V}_2$ matrix in Eq.~\eqref{eqn:YPR} is a flat direction of the flavon potential, as are $U$, $V$ and $\hat V$. The quark mixing matrix of the full theory, however, is no longer a flat direction: It is lifted by the $1+2$ flavor-locked structure to an $\mathcal{O}(1)$ $2 \oplus 1$ block form with all other entries suppressed by small ratios of quark masses. Diagonalizing the quark mass matrices resulting from (\ref{eqn:PPMM}), one finds the following schematic predictions for the CKM matrix elements
\renewcommand*{\arraystretch}{.7}
\begin{equation}
\label{eqn:GSCKM}
 \mathcal{V}_{\text{ckm}} \sim \begin{pmatrix} 1 & \mathcal{O}(\theta) & \mathcal{O}(m_d/m_b) \\ \mathcal{O}(\theta) & 1 & \mathcal{O}(m_s/m_b) \\ \mathcal{O}(m_d/m_b) & \mathcal{O}(m_s/m_b) & 1 \end{pmatrix} ~,
\end{equation}
where $\theta$ is the rotation angle in the $\mathcal{V}_2$ matrix (see Eq.~\eqref{eqn:PPMM}), that is a priori a free parameter of $\mathcal{O}(1)$.
This structure suggests that the observed CKM hierarchies can be accommodated: The $1$--$3$ and $2$--$3$ mixing elements are automatically suppressed at a level that resembles the experimental values.

In the decoupling/alignment limit $\cos(\beta-\alpha) = 0$, flavor-violating processes from heavy Higgs exchange vanish in the large $m_{H,A}$ limit. However, from Eqs.~\eqref{eqn:PPMM} and~\eqref{eqn:GSCKM} it is not obvious whether the flavor structure of the $1+2$ flavor-locked configuration reduces to the SM in an appropriate limit. As a demonstration that the $1+2$ flavor-locked configuration is compatible with data, we  heuristically identified the following example input parameters,
\begin{subequations}
\label{eq:benchmark}
\begin{align} 
&  z_t \frac{r}{\Lambda_F} \frac{v_1}{\sqrt{2}} \simeq 173~\text{GeV} \,, && z_c \frac{r}{\Lambda_F} \frac{v_2}{\sqrt{2}} \simeq 1.9~\text{GeV} \,, && z_u \frac{r}{\Lambda_F} \frac{v_2}{\sqrt{2}} \simeq 7~\text{MeV} ~, \nn \\
&  z_b \frac{\hat r}{\Lambda_F} \frac{v_1}{\sqrt{2}} \simeq 4.8~\text{GeV} \,, && z_s \frac{\hat r}{\Lambda_F} \frac{v_2}{\sqrt{2}} \simeq 240~\text{MeV} \,, && z_d \frac{\hat r}{\Lambda_F} \frac{v_2}{\sqrt{2}} \simeq 21~\text{MeV} ~,
\end{align}
\begin{align} 
 & \theta_{13} \simeq -0.2 \,, && \theta_{23} \simeq -0.1 \,, && \vartheta_{13} \simeq 1.0 \,, && \vartheta_{23} \simeq 1.0 \,, && \hat\vartheta_{13} \simeq 0.4 \,, && \hat\vartheta_{23} \simeq 1.5 \,, \nn \\
 & \theta \simeq 0.1 \,, && \psi_d \simeq -2.1 \,, &&\psi_s \simeq -0.2 \,,
\end{align}
\end{subequations}
and $\psi_u=\psi_c=\psi_m = 0$, where we have defined the two vevs, $v_1\equiv v\cos\beta$ and $v_2\equiv v\sin\beta$. The phases $\psi_u,\psi_c,\psi_m$ are set to zero for simplicity, as they have negligible impact on all the observables that we are considering. (The phases $\psi_u,\psi_c$ enter in $D^0$--$\bar D^0$  mixing, but, as we will discuss in Sec. \ref{sec:MesonMix}, they are only very weakly constrained.) 
This parameter set leads to the theoretical predictions shown in Table~\ref{tab:tree} for the six quark masses and a set of five CKM elements. 

We compare these predictions to data for the quark masses and CKM parameters, shown in Table~\ref{tab:tree}. To be self-consistent, we use data only from processes that are insensitive to heavy Higgs 
exchange, i.e. processes that are tree-level in the SM. (Since we are ultimately interested in considering the phenomenology of collider-accessible heavy Higgs bosons, loop-level processes in the SM will receive 
corrections from heavy Higgs exchanges, but measurements of tree-level processes will be insensitive to these effects.) To reproduce the Cabibbo angle 
$\lambda_C \simeq 0.22506 \pm 0.00050$~\cite{Patrignani:2016xqp}, $\theta$ needs to be constrained accordingly to a narrow $\mathcal{O}(1)$ range. Since we require only a mixing matrix with canonical entries 
of the same characteristic size as observed in Nature, we do not insist on such a narrow range for $\theta$. Similarly, for comparison of the theoretical predictions to data, instead of using the experimental
uncertainties of the observables (which in some cases are measured with remarkable precision), we choose $10\%$ uncertainties for all CKM parameters and the bottom, charm, and strange masses. 
In the case of the top mass we chose a 1.5~GeV uncertainty, while for the up and down masses we use 100\% uncertainties.   
Using these values, the theoretical predictions for the benchmark point~\eqref{eq:benchmark} are in excellent agreement with the observed quark masses and CKM parameters. 

\renewcommand{\arraystretch}{1.1}
\newcolumntype{C}{ >{\centering\arraybackslash $} m{3cm} <{$}}
\newcolumntype{D}{ >{\raggedright\arraybackslash $} c <{\qquad $}}
\begin{table}[tb]
\begin{tabular}{DCC|DCC}
\hline
\hline
 & \text{Mass Data} & \text{Benchmark} & & \text{CKM Data} & \text{Benchmark} \\
\hline
\hline
m_t 	& 173.5 \pm 1.5~\text{GeV} 	& \simeq 173~\text{GeV} 		& |V_{us}| 	& \multirow{2}{*}{$0.225 \pm 0.023$} &  \multirow{2}{*}{$\simeq  0.23$} \\
m_b 	& 4.8 \pm 0.5~\text{GeV} 		& \simeq 4.8~\text{GeV} 		& |V_{cd}| &&\\
m_c 	& 1.7 \pm 0.2~\text{GeV} 		& \simeq 1.7~\text{GeV} 		& |V_{cb}| 	& (40.5 \pm 4.1) \times 10^{-3} 		& \simeq  40 \times 10^{-3} \\
m_s 	& 100 \pm 10~\text{MeV} 		& \simeq 100~\text{MeV}		& |V_{ub}| 	& (4.1 \pm 0.4) \times 10^{-3} 		& \simeq  4.1\times 10^{-3} \\
m_u 	& 2.0 \pm 2.0~\text{MeV} 		& \simeq 2~\text{MeV} 		&  \gamma 	& 73.2 \pm 7.3^\circ 				& \simeq  71^\circ \\
m_d & 5.0 \pm 5.0~\text{MeV}		& \simeq 5~\text{MeV}		& &&\\
\hline
\hline
\end{tabular}
\caption{\label{tab:tree} Data for quark (pole) masses and CKM parameters used in our analysis. The central values correspond to the measured quark masses~\cite{Patrignani:2016xqp} and 
CKM parameters~\cite{Hocker:2001xe, Charles:2004jd}. All CKM parameters and the $b$, $c$, and $s$ quark masses are assigned $10$\% uncertainties. In the case of the top mass we use a 1.5~GeV 
uncertainty, while for the up and down masses we use 100\% uncertainties. Also shown are predictions corresponding to the benchmark point~\eqref{eq:benchmark}.}
\end{table}

To quantify the ``goodness'' of the benchmark or other points in the parameter space, we construct a $\chi^2$-like function, $X_{\text{tree}}^2$, for the six quark masses and CKM elements measured from tree-level processes,
\begin{equation}\label{eq:chi2}
 X^2_\text{tree} = \sum_{i = u,c,t,d,s,b} \Bigg[\frac{(m_i^\text{FL}-m_i)^2}{(\sigma_{m_i})^2}\Bigg] + \sum_{i = us, cd, cb, ub} \Bigg[\frac{(|V_i|^\text{FL}-|V_i|)^2}{(\sigma_{V_i})^2}\Bigg]  + \frac{(\gamma^\text{FL}-\gamma)^2}{(\sigma_{\gamma})^2} \,.
\end{equation}
where the `$\text{FL}$' superscript denotes the theory prediction at a given point in the flavor-locked theory parameter space~\eqref{eqn:PPMM}, and we treat the uncertainties as uncorrelated.  
While such a $X^2$ function implies a well-defined $p$-value for a goodness-of-fit of the quoted data to a given theory point, one cannot construct from $X^2$ a sense of the probability for a given 
theory to produce the observed flavor data and hierarchies. Instead, the $X^2$ function allows us only to understand whether or not the flavor-locked configuration results generically in a flavor structure that
agrees with observation at the level of tens of percent.

\begin{figure}[ht!]
\includegraphics[width=0.46\textwidth]{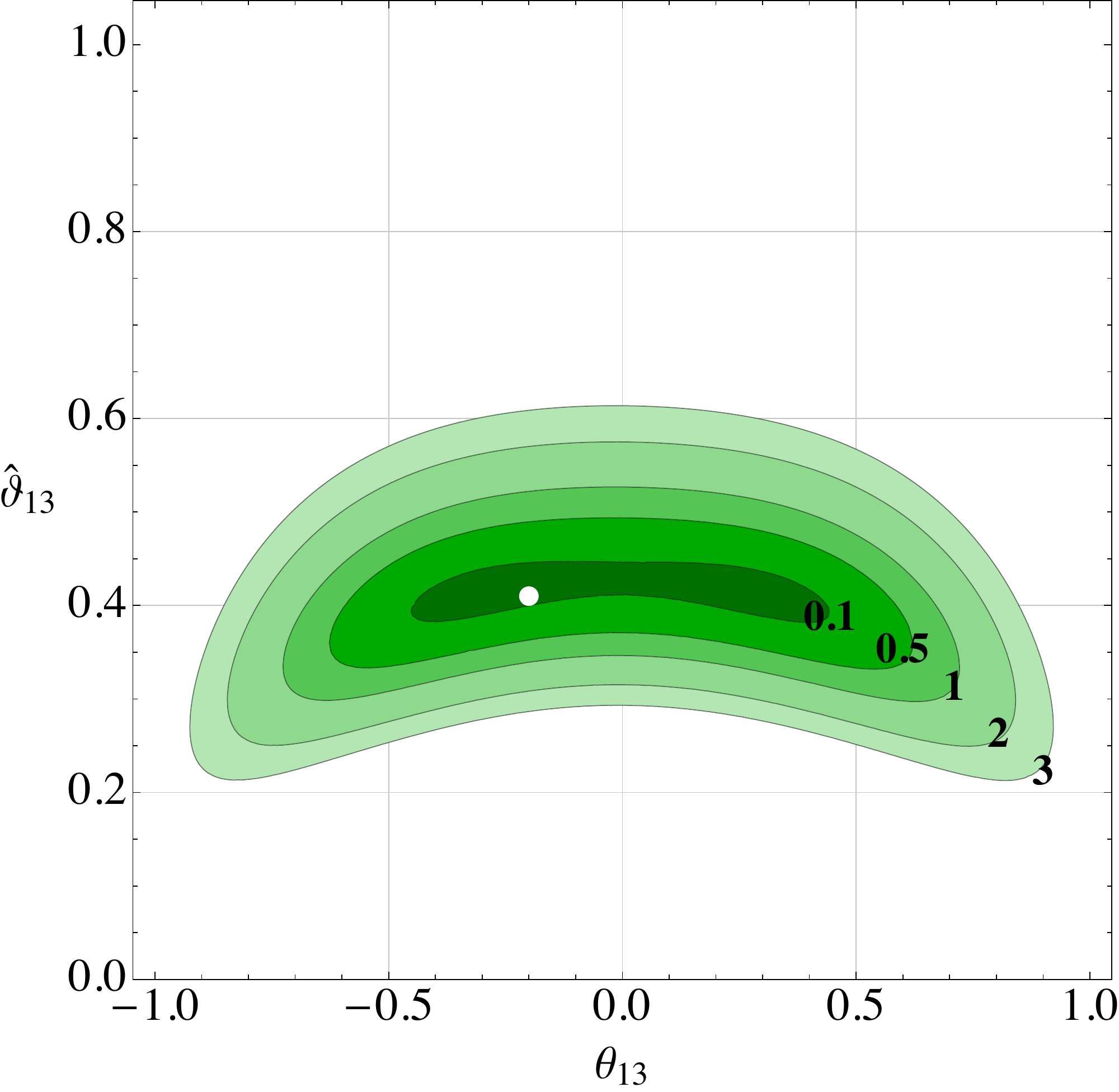}~~~
\includegraphics[width=0.46\textwidth]{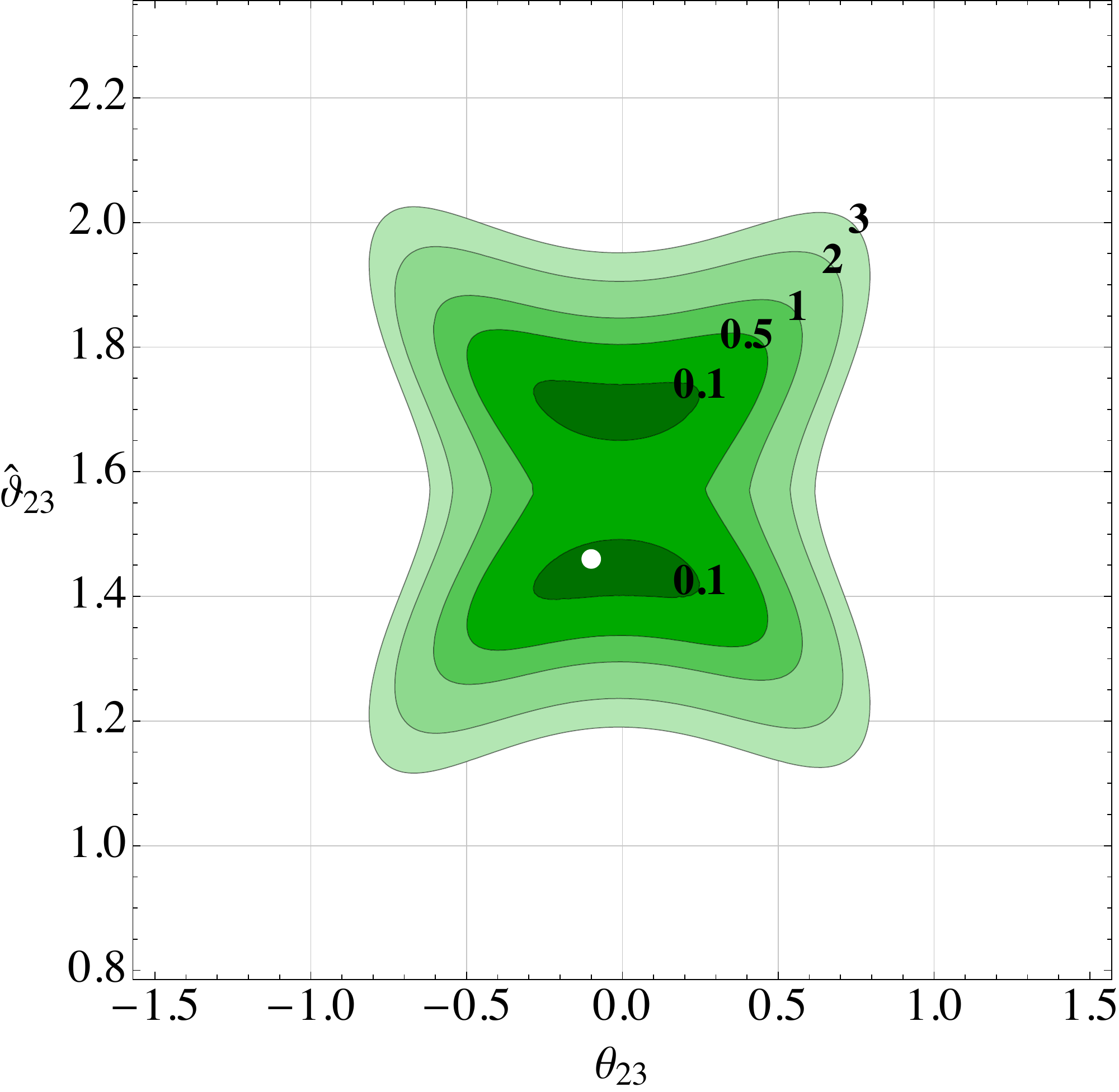}\\[12pt]
\includegraphics[width=0.46\textwidth]{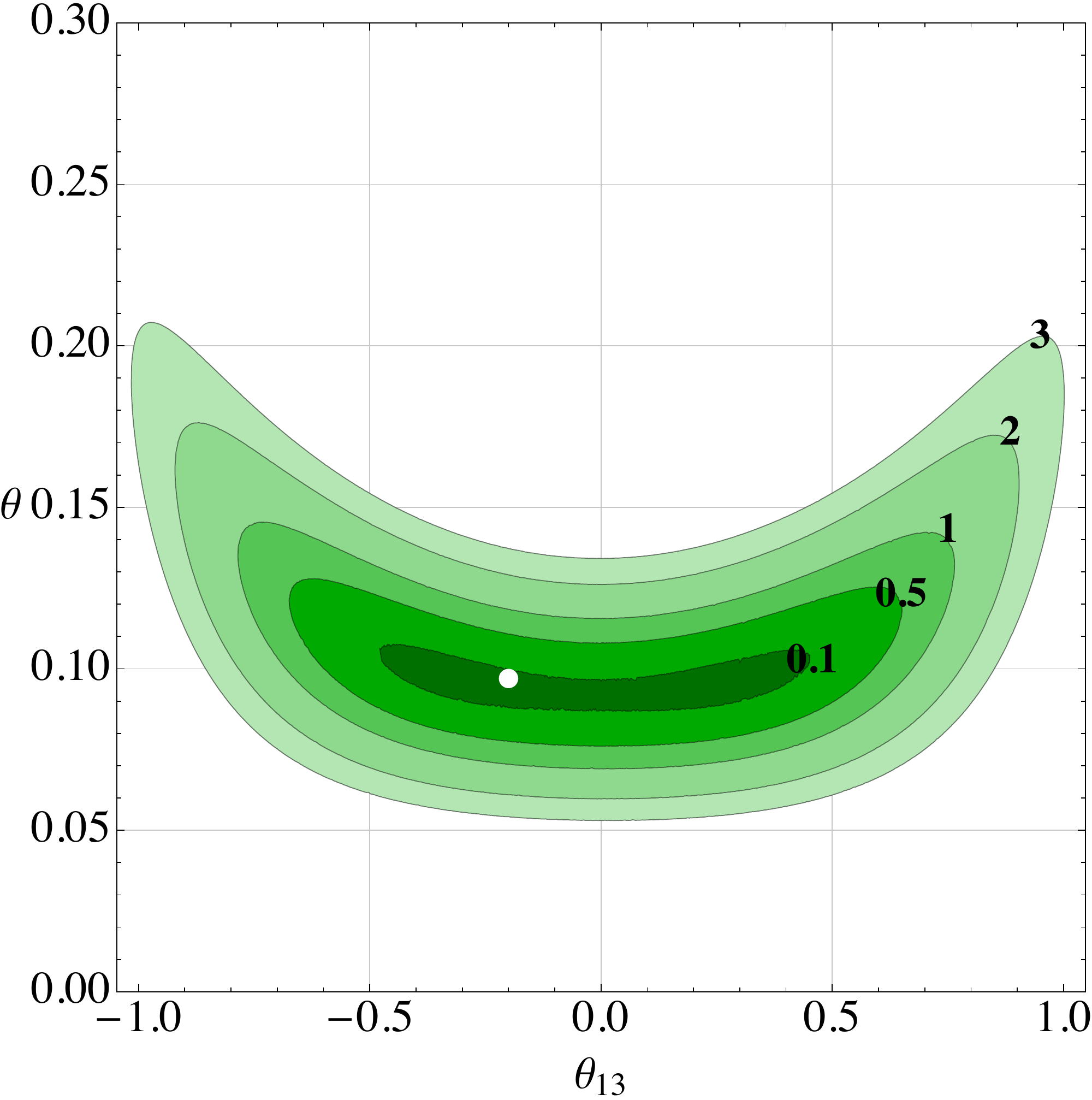}~~~
\includegraphics[width=0.46\textwidth]{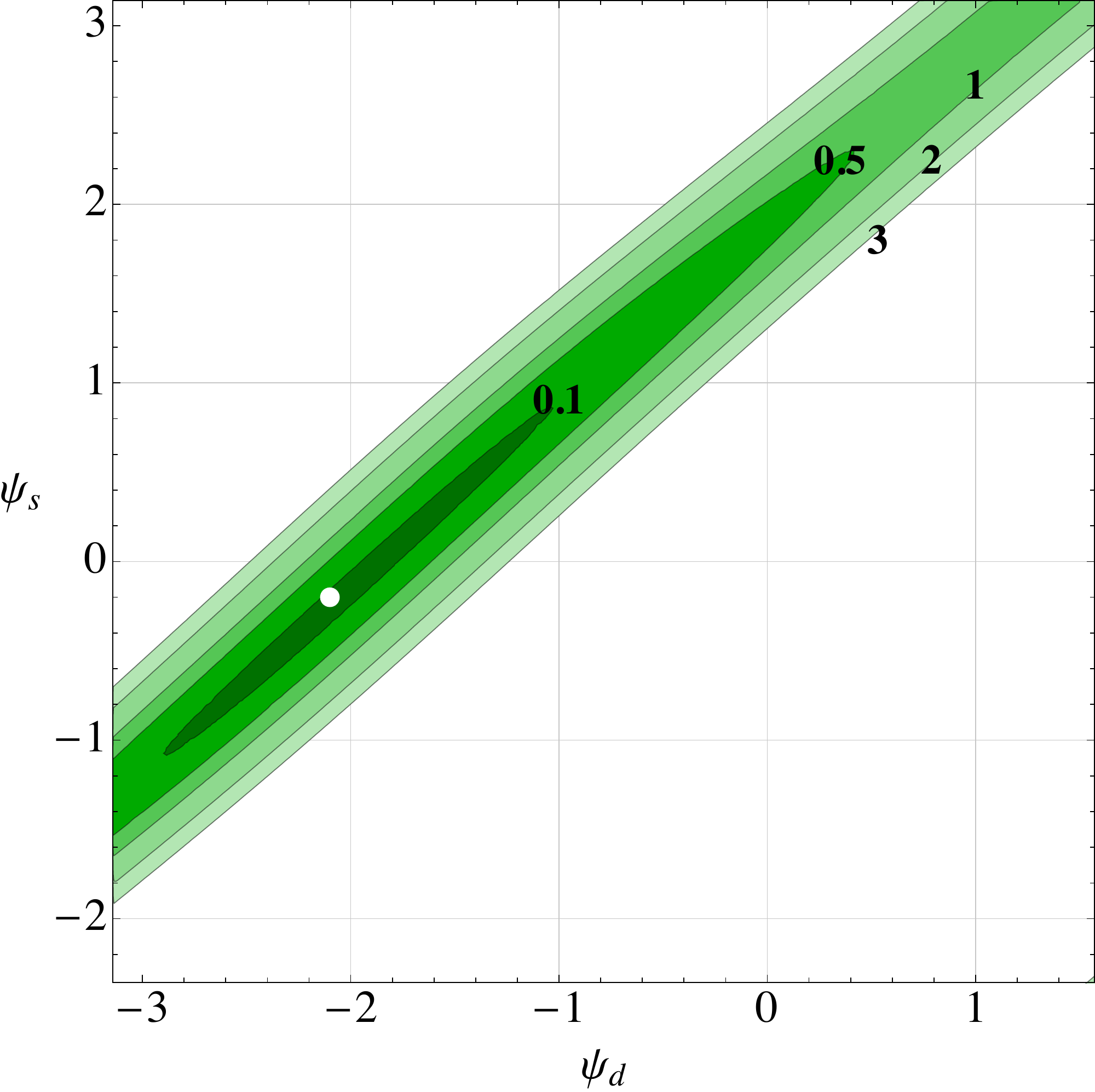}
\caption{$X^2_{\text{tree}}/\text{dof}$ regions on various two-dimensional slices of the $1+2$ flavor-locked theory parameter space in the neighborhood of the benchmark point~\eqref{eq:benchmark}. 
Contour values are labeled in black; the benchmark point~\eqref{eq:benchmark} is shown by the white circle.}
\label{fig:CKM}
\end{figure}

In Fig.~\ref{fig:CKM} we show the $X^2_{\text{tree}}$ behavior of the flavor model on various two-dimensional parametric slices in the neighborhood of the benchmark point~\eqref{eq:benchmark}, which is 
denoted by the white circle. That is, in each plot, all the theory parameters are fixed to the benchmark values in Eqs.~\eqref{eq:benchmark}, except for the two parameters corresponding to the 
plot axes. The number of degrees of freedom (dof) in the $X^2_{\text{tree}}$ statistic is then $11-2=9$. The contours show regions of $X^2_{\text{tree}}/\text{dof}$ that lead to an overall good agreement 
between the observed quark masses and CKM parameters and those predicted in the model.

As can be seen from the plots in Fig.~\ref{fig:CKM}, there are extended regions of parameter space where there is fairly good agreement between the theory predictions and the measured quark masses 
and CKM parameters. In particular, $\mathcal{O}(1)$ variations of the mixing angles $\theta_{13}, \theta_{23}, \vartheta_{13}, \vartheta_{23}, \hat\vartheta_{13}, \hat\vartheta_{23}$ around the benchmark 
point are possible, without worsening the agreement substantially. Only the angle $\theta$ that sets the Cabibbo angle is strongly constrained and has to be set to a narrow range by hand. This behavior 
should be contrasted to the SM, for which two CKM mixing angles -- i.e. the suppressed $1$--$3$ and $2$--$3$ mixings -- have to be tuned small.  

\subsection{Constraints from meson mixing}\label{sec:MesonMix}
As mentioned above and in Sec.~\ref{sec:F2HDM}, the neutral Higgs bosons of the F2HDM setup generically have flavor violating couplings. In particular, their tree-level exchange will contribute to meson oscillations. For kaon oscillations the corresponding new physics (NP) contribution to the mixing amplitude is given by
\begin{multline} \label{eq:M12K}
 M_{12}^\text{NP} = m_K^3 \frac{f_K^2}{v^2} \frac{1}{s_\beta^2 c_\beta^2} \Bigg[ \frac{1}{4} B_4 \eta_4 \bigg( \frac{c_{\beta-\alpha}^2}{m_h^2} + \frac{s_{\beta-\alpha}^2}{m_H^2} + \frac{1}{m_A^2} \bigg) \frac{m_{sd}^\prime {}^* m_{ds}^\prime}{m_s^2} \nn \\
  - \bigg(\frac{5}{48} B_2 \eta_2 - \frac{1}{48} B_3 \eta_3 \bigg) \bigg( \frac{c_{\beta-\alpha}^2}{m_h^2} + \frac{s_{\beta-\alpha}^2}{m_H^2} - \frac{1}{m_A^2} \bigg) \frac{(m_{sd}^\prime {}^*)^2 + (m_{ds}^\prime)^2}{m_s^2} \Bigg] ~.
\end{multline}
The $m^\prime$ parameters are the off-diagonal entries of the contribution to the down quark mass matrix from the $H_2$ doublet in the quark mass eigenstate basis, and are fully determined by the parameters entering the $1+2$ flavor-locked Yukawas~\eqref{eqn:PPMM}.
The NP mixing amplitude also depends on the heavy Higgs masses $m_H$ and $m_A$, the ratio of the two Higgs vacuum expectation values $\tan\beta$ and the scalar mixing angle $\alpha$. 
As additional parametric input in Eq.~\eqref{eq:M12K}, we have the kaon decay constant $f_K \simeq 155.4$~MeV~\cite{Dowdall:2013rya}.
The bag parameters  $B_2 \simeq 0.46$, $B_3 \simeq 0.79$, $B_4 \simeq 0.78$ are evaluated at the scale $\mu_K = 3$~GeV and are taken from Ref.~\cite{Carrasco:2015pra} (see also Refs.~\cite{Jang:2015sla,Garron:2016mva}). The parameters $\eta_i$ encode renormalization group running effects. From 1-loop RGEs we find
\begin{align}
\eta_2 &\simeq 0.68\,,&
\eta_3 &\simeq -0.03\,,&
\eta_4 &= 1\,.
\end{align}

The relevant observables that are measured in the neutral kaon system are the mass
difference $\Delta M_K$ and the CP violating parameter $\epsilon_K$. The experimental results and the corresponding SM predictions and uncertainties are collected in Table~\ref{tab:input}. In terms of the NP mixing amplitude, these observables are given by
\begin{equation}
 \Delta M_K = \Delta M_K^\text{SM} + 2 \text{Re}(M_{12}^\text{NP}) ~, ~~~~~ \epsilon_K = \epsilon_K^\text{SM} + \kappa_\epsilon \frac{\text{Im}(M_{12}^\text{NP})}{\sqrt{2} \Delta M_K}~.
\end{equation}
In the expression for $\epsilon_K$ we use $\kappa_\epsilon = 0.94$~\cite{Buras:2010pza} and the measured value of $\Delta M_K$ shown in Table~\ref{tab:input}.

\renewcommand{\arraystretch}{1.3}
\newcommand{\tcite}[1]{\text{\cite{#1}}}
\begin{table}[tb]
\newcolumntype{C}{ >{\centering\arraybackslash $} m{5cm} <{$}}
\newcolumntype{D}{ >{\raggedright\arraybackslash $} c <{\qquad $}}
\newcolumntype{E}{ >{\centering\arraybackslash $} m{3cm} <{$}}
\begin{tabular}{D|CCE}
\hline
\hline
 			& \text{Data} 		& \text{SM Prediction}		 & \text{NP Contribution}\\
\hline
\hline
\Delta M_K 		&(5.294 \pm 0.002)\times 10^{-3}~\text{ps}^{-1} ~\tcite{Patrignani:2016xqp} 	& (4.7 \pm 1.8) \times 10^{-3}~\text{ps}^{-1}  ~\tcite{Brod:2011ty} 	& \simeq -2 \times 10^{-6}~\text{ps}^{-1}\\
\Delta M_{B_d} 		&0.5055 \pm 0.0020~\text{ps}^{-1} ~\tcite{Amhis:2016xyh} 				& 0.63 \pm 0.07 ~\text{ps}^{-1} ~\tcite{Bazavov:2016nty} 			& \simeq 0.01~\text{ps}^{-1}\\
\Delta M_{B_s} 		&17.757 \pm 0.021~\text{ps}^{-1} ~\tcite{Amhis:2016xyh} 					&19.6 \pm 1.3  ~\text{ps}^{-1} ~\tcite{Bazavov:2016nty} 			& \simeq -1.8~\text{ps}^{-1}\\
\epsilon_K 		& (2.288 \pm 0.011) \times 10^{-3}  ~\tcite{Patrignani:2016xqp} 				&  (1.81\pm 0.28) \times 10^{-3}  ~\tcite{Brod:2011ty} 			& \simeq 0.025 \times 10^{-3}\\
\phi_d 			& 43.7 \pm 2.4^\circ  ~\tcite{Charles:2004jd} 							&  47.5 \pm 2.0^\circ ~\tcite{Charles:2004jd}  					& \simeq  -2.4^\circ\\
\phi_s 			& -1.2 \pm 1.8^\circ ~\tcite{Amhis:2016xyh} 							& -2.12 \pm 0.04^\circ ~\tcite{Charles:2004jd} 					& \simeq  0.26^\circ\\
\hline
\hline
\end{tabular}
\caption{\label{tab:input} Experimental measurements and SM predictions for meson mixing observables. The SM prediction for $\Delta M_K$ and its uncertainty refers to the short distance contribution. To account for long distance effects, we  use $\Delta M_K^\text{SM} = \Delta M_K^\text{exp} ( 1 \pm 0.5 )$ in our numerical analysis. Also shown are the NP contributions corresponding to the benchmark point~\eqref{eq:benchmark}.}
\end{table}

In the case of neutral $B$ meson oscillations, we find it convenient to normalize the NP mixing amplitude directly to the SM amplitude. For $B_s$ mixing we find
\begin{multline}\label{eq:M12B}
 \frac{M_{12}^\text{NP}}{M_{12}^\text{SM}} 
 = \frac{m_{B_s}^2}{s_\beta^2 c_\beta^2} \frac{16\pi^2}{g_2^2} \frac{1}{S_0} \Bigg[ 2 \xi_4 \bigg( \frac{c_{\beta-\alpha}^2}{m_h^2} + \frac{s_{\beta-\alpha}^2}{m_H^2} + \frac{1}{m_A^2} \bigg) \frac{m_{bs}^\prime {}^* m_{sb}^\prime}{ m_b^2 (V_{tb} V_{ts}^*)^2} \\
   + \big(\xi_2 + \xi_3 \big) \bigg( \frac{c_{\beta-\alpha}^2}{m_h^2} + \frac{s_{\beta-\alpha}^2}{m_H^2} - \frac{1}{m_A^2} \bigg) \frac{(m_{bs}^\prime {}^*)^2 + (m_{sb}^\prime)^2}{ m_b^2 (V_{tb} V_{ts}^*)^2 } \Bigg] ~.
\end{multline}
A completely analogous expression holds for $B_d$ oscillations. The SM loop function $S_0 \simeq 2.3$, and
the $\xi_i$ factors contain QCD running as well as ratios of hadronic matrix elements.
At 1-loop we find
\begin{align}
\xi_2 & \simeq -0.47 ~(-0.47)\,, &
\xi_3 & \simeq -0.005 ~(-0.005)\,,&
\xi_4 & \simeq 0.99 ~(1.03)\,,
\end{align}
where the first (second) value corresponds to $B_s$ ($B_d$) mixing. To obtain these values we used bag parameters from Ref.~\cite{Carrasco:2013zta} (see also Ref.~\cite{Bazavov:2016nty}). The meson oscillation frequencies and the phases of the mixing amplitudes are given by
\begin{align}
 \Delta M_s & = \Delta M_s^\text{SM} \times \bigg| 1 + \frac{M_{12}^\text{NP}}{M_{12}^\text{SM}} \bigg|\,, &  \phi_s & = -2\beta_s + \text{Arg}\bigg( 1 + \frac{M_{12}^\text{NP}}{M_{12}^\text{SM}}\bigg) ~, \\
 \Delta M_d & = \Delta M_d^\text{SM} \times \bigg| 1 + \frac{M_{12}^\text{NP}}{M_{12}^\text{SM}} \bigg|\,, & \phi_d & = 2\beta + \text{Arg}\bigg( 1 + \frac{M_{12}^\text{NP}}{M_{12}^\text{SM}} \bigg) ~.
\end{align}
The experimental results and the corresponding SM predictions and uncertainties for the observables are collected in Table~\ref{tab:input}.
Note that the NP contributions to the kaon and B meson mixing amplitudes~\eqref{eq:M12K} and~\eqref{eq:M12B} vanish in the decoupling limit $\cos(\beta - \alpha) = 0$, $m_A, m_H \to \infty$. The 
NP effects in $D^0$--$\bar D^0$ oscillations are suppressed by the tiny up quark mass. We have explicitly checked that $D^0$--$\bar D^0$  oscillations do not lead to relevant constraints. 

In the case that the heavy Higgs masses are below the TeV scale, the NP effects in the mixing observables do not vanish, and we proceed to investigate the size of such effects. For the following 
numerical study, we will set the heavy Higgs masses to a benchmark value, $m_H = m_A = 500~$GeV. We use a moderate value of $\tan\beta = 5$, and work in the alignment limit $\beta-\alpha=\pi/2$. 
For the benchmark parameters in Eq.~\eqref{eq:benchmark}, we show the NP contributions to meson mixing observables in the last column of Table~\ref{tab:input}.
For the benchmark point, the NP contributions are in most cases within the combined experimental and SM uncertainties.

Similar to Eq.~\eqref{eq:chi2}, we construct a $X^2_\text{loop}$ function, that compares the NP contributions to the difference of the data and SM predictions, for  the three mass differences $\Delta M_K$, $\Delta M_d$, and $\Delta M_s$, as well as the CP violating observables $\epsilon_K$, $\phi_d$, and $\phi_s$. That is,
\begin{equation}
 X^2_\text{loop} = \sum_{i = K,d,s}\Bigg[ \frac{(\Delta M_i^\text{NP}-\Delta M_i^\text{exp-SM})^2}{(\sigma_{\Delta M_i^\text{exp}})^2+(\sigma_{\Delta M_i^\text{SM}})^2}\Bigg] + \sum_{i = d,s} \Bigg[\frac{(\phi_i^\text{NP}-\phi_i^\text{exp-SM})^2}{(\sigma_{\phi_i^\text{exp}})^2+(\sigma_{\phi_i^\text{SM}})^2}\Bigg] + \frac{(\epsilon_K^\text{NP}-\epsilon_K^\text{exp-SM})^2}{(\sigma_{\epsilon_K^\text{exp}})^2+(\sigma_{\epsilon_K^\text{SM}})^2} ~,
\end{equation}
where the superscript `exp-SM' indicates that we are using the difference of the measured values and the SM predictions given in Table~\ref{tab:input}. 

\begin{figure}[ht!]
\includegraphics[width=0.46\textwidth]{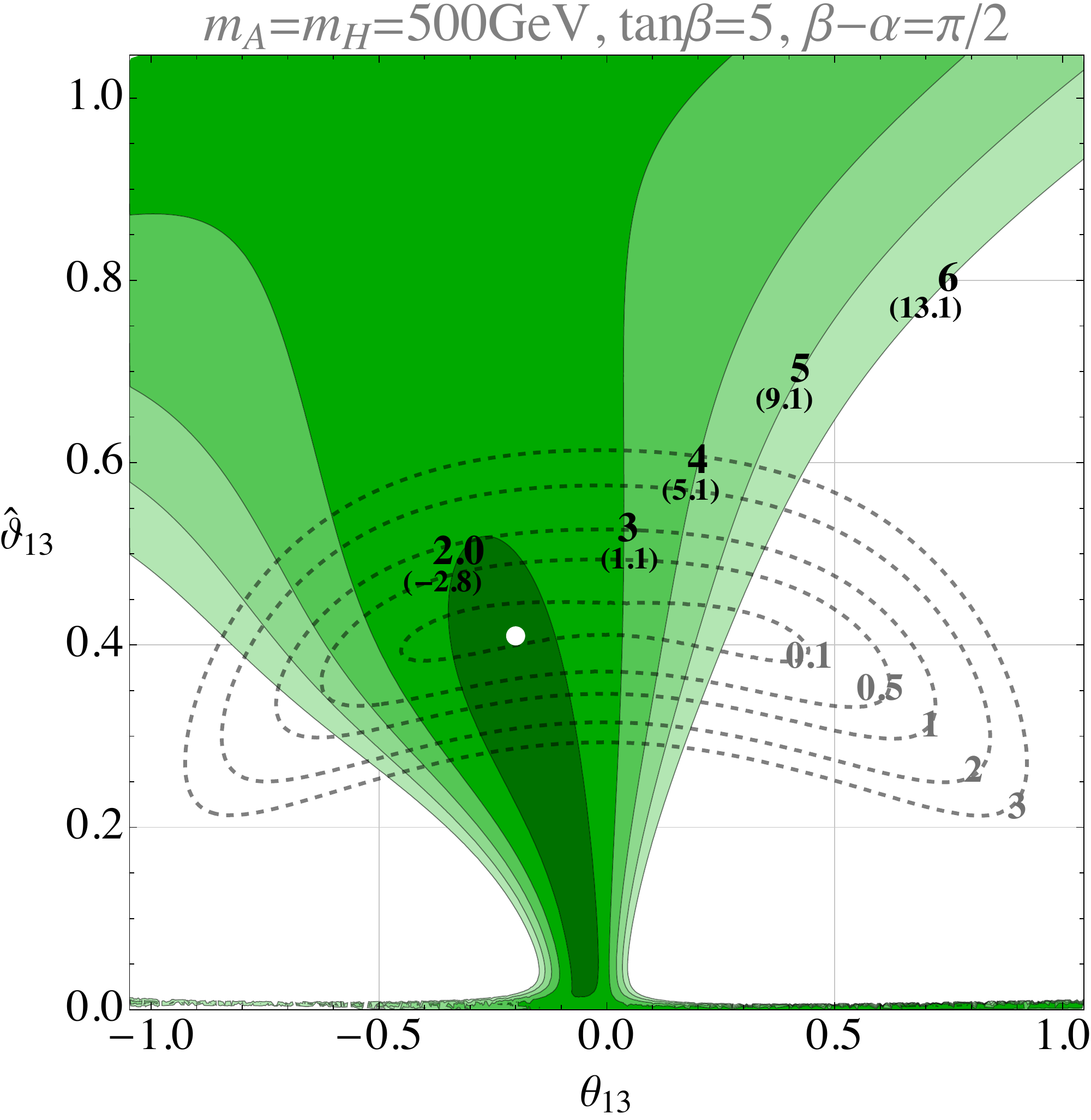}~~~
\includegraphics[width=0.46\textwidth]{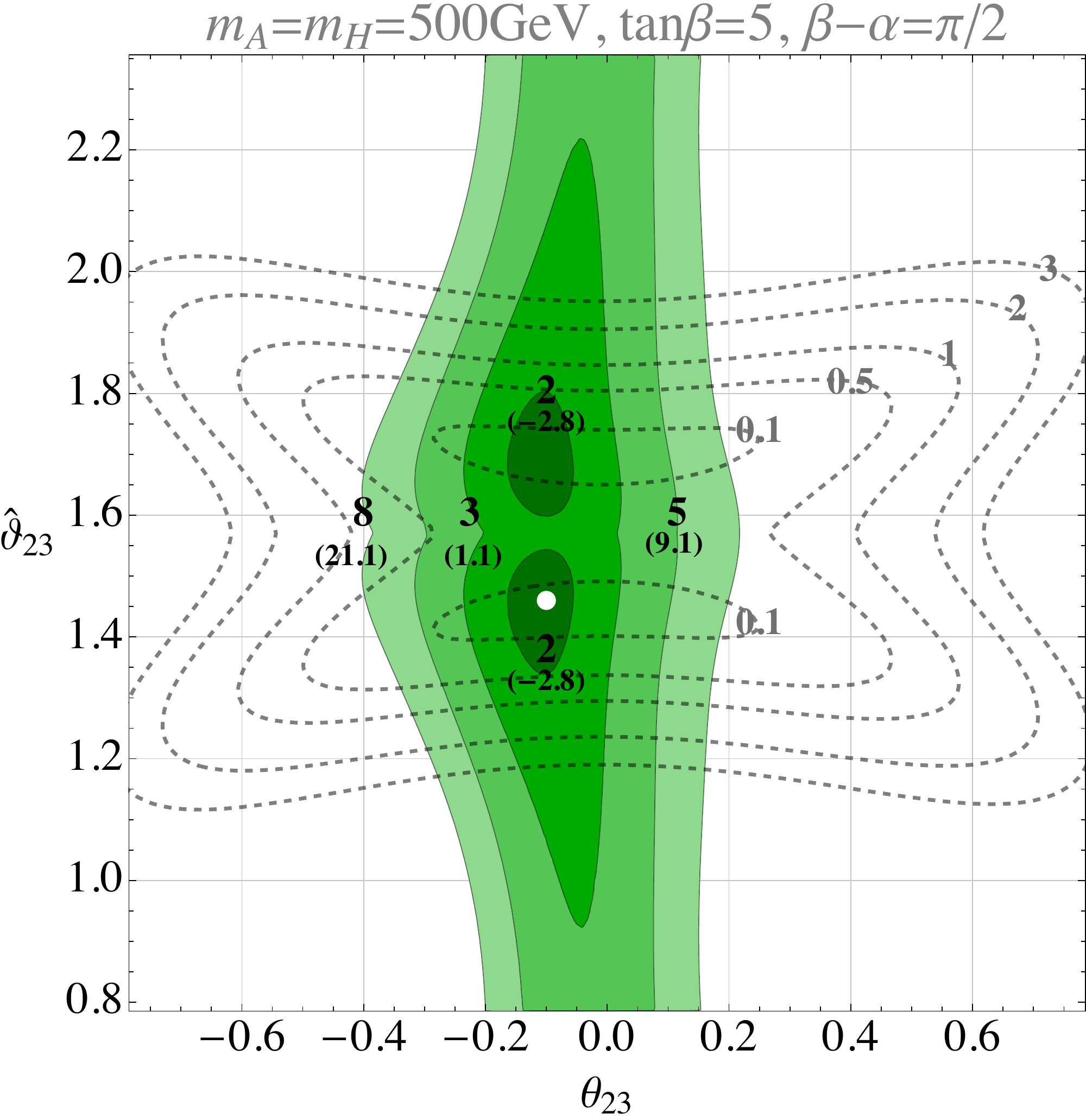}\\[12pt]
\includegraphics[width=0.46\textwidth]{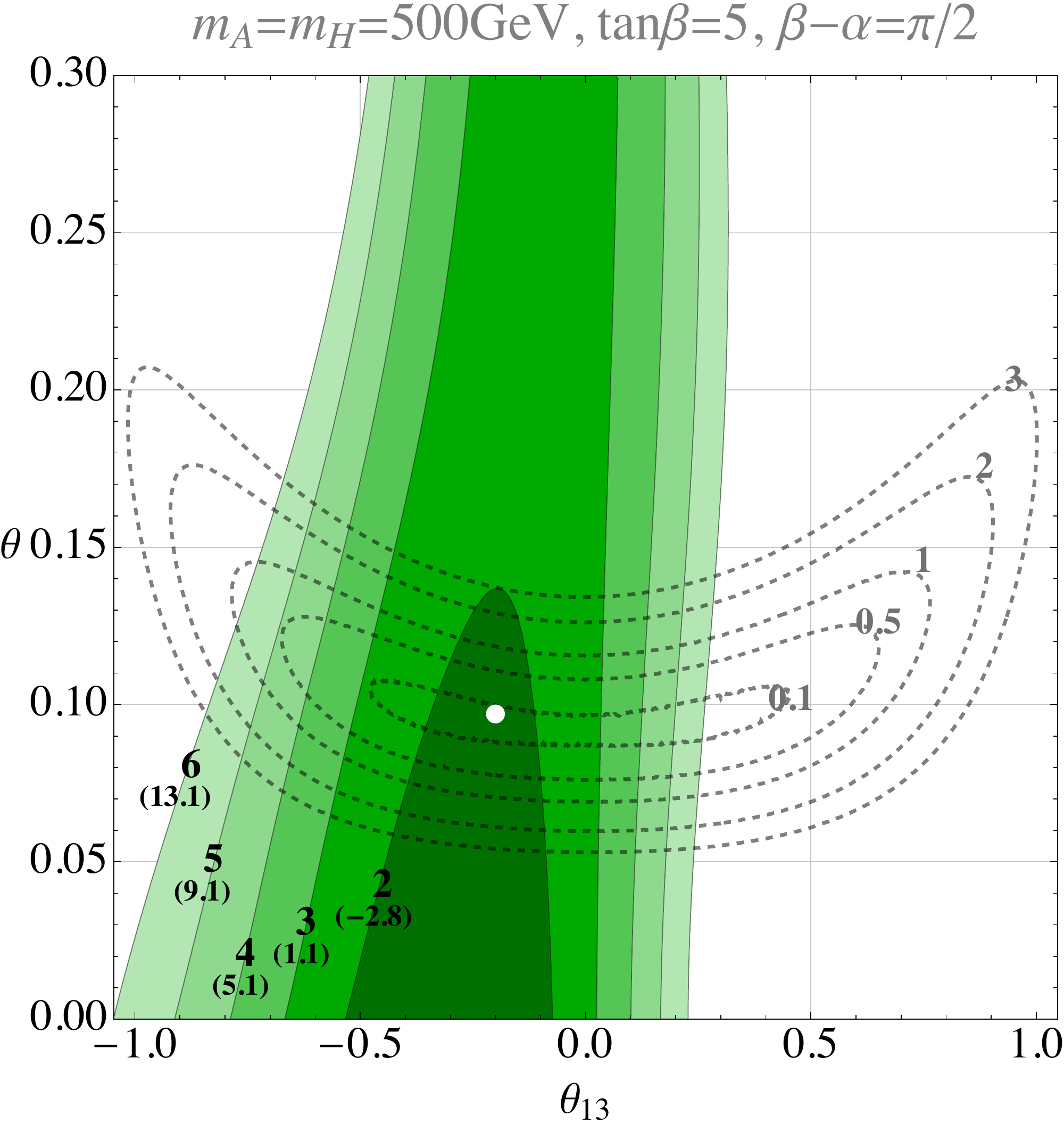}~~~
\includegraphics[width=0.46\textwidth]{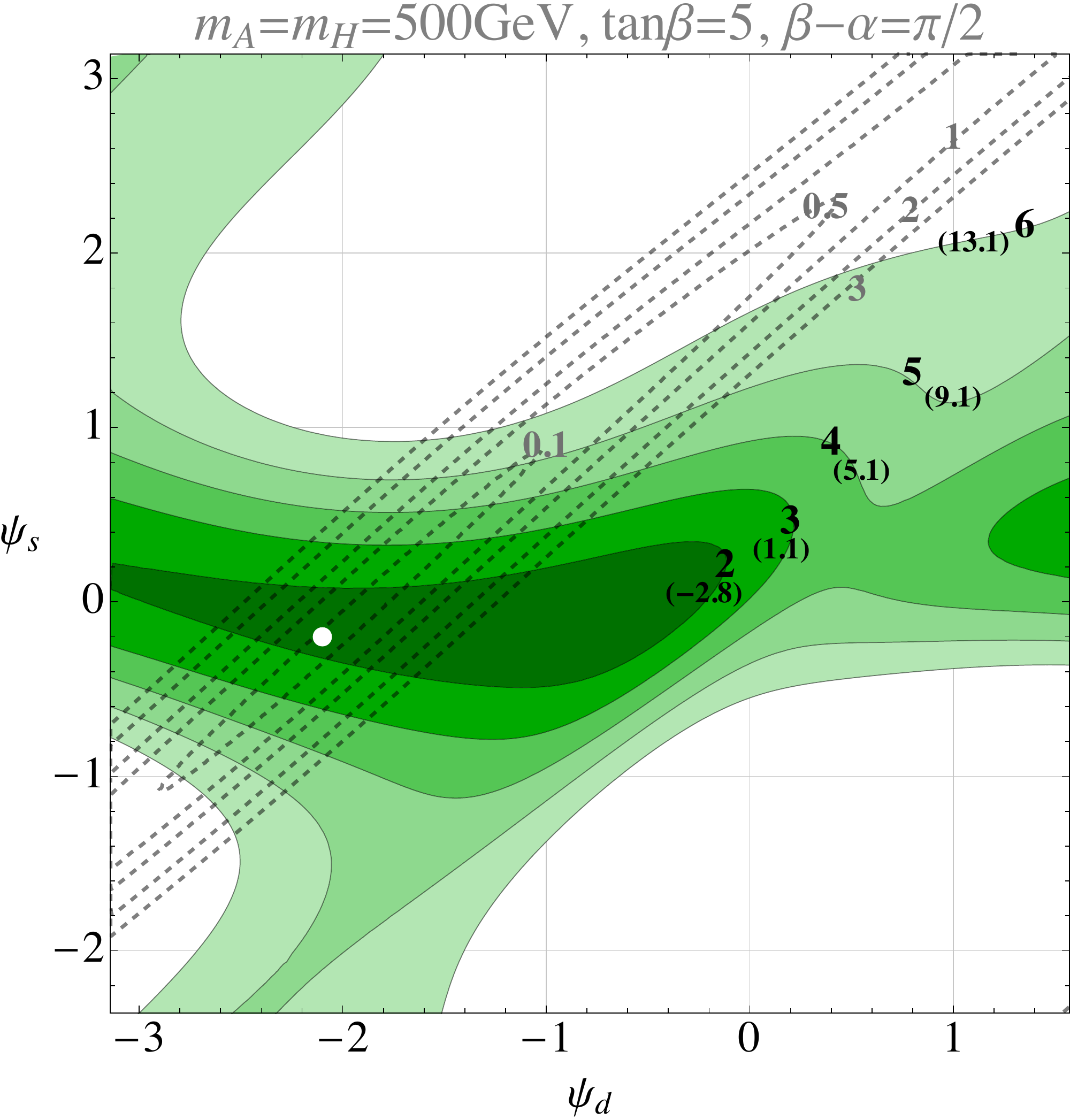}
\caption{$X^2_{\text{loop}}/\text{dof}$ regions on various two-dimensional slices of the $1+2$ flavor-locked theory parameter space in the neighborhood of the benchmark point~\eqref{eq:benchmark}. 
Contour values are labeled in black; we also show the values for $X^2_\text{loop} - X^2_\text{loop}(\text{SM})$ in parentheses. The benchmark point~\eqref{eq:benchmark} is shown by the white circle. The contours from Fig.~\ref{fig:CKM} are shown by the dotted lines with the corresponding contours labeled in gray.}
\label{fig:mixing}
\end{figure}

Fig.~\ref{fig:mixing} shows the $X^2_{\text{loop}}/\text{dof}$ behavior of the flavor model on various two-dimensional parametric slices in the neighborhood of the benchmark point~\eqref{eq:benchmark}.
As for Fig.~\ref{fig:CKM}, on each slice all theory parameters are fixed to the benchmark values~\eqref{eq:benchmark}, except for the two parameters corresponding to the plot axes. 
The number of degrees of freedom in the $X^2_{\text{loop}}$ statistic is then $6-2=4$.
Note that the SM predictions and experimental results for meson mixing observables from Table~\ref{tab:input} show slight tensions~\cite{Bazavov:2016nty,Blanke:2016bhf,Buras:2016dxz}, as indicated by 
the non-negligible SM contribution to the $X^2_\text{loop}$ function, $X^2_\text{loop}(\text{SM}) \simeq 10.8$.  We observe that ranges of model parameters exist for which $X^2$ is mildly better than in the SM: At our benchmark $X^2_\text{loop} - X^2_\text{loop}(\text{SM})\simeq -3.7$. (Identifying all regions of parameter space of our framework that can address existing tensions 
in meson observables is left for future studies.) Moreover, comparing with the contours obtained from the $X^2_\text{tree}/\text{dof}$ function (dotted lines), 
we find that extended regions of parameter space exist where CKM elements and masses as well as meson mixing observables are described in a satisfactory way. 

\section{Conclusion and outlook} \label{sec:conclusions}

We have presented a new framework to address the SM flavor puzzle, synthesizing the structure of the `flavorful' 2HDM with the `flavor-locking' mechanism. This mechanism makes use of distinct flavon and hierarchon 
sectors to dynamically generate arbitrary quark mass hierarchies, without assigning additional symmetries to the quark fields themselves. In this paper, we have shown that with suitable symmetry assignments 
in the flavon and hierarchon sectors, the global minimum of the general renormalizable flavon potential can be identified with a `flavor-locked' configuration: An aligned, rank-1 configuration for each flavon, and 
arbitrary (block) unitary misalignment between the up and down quark Yukawas, so that a unique hierarchon vev controls each quark mass.

In the presence of only one SM-like Higgs doublet, this leads to quark mixing angles that are generically $\mathcal{O}(1)$. Introducing instead a flavorful 2HDM Higgs sector -- two Higgs doublets, such that one 
Higgs couples only to the third generation, while the other couples to the first two generations -- leads to a $1+2$ flavor-locked theory. We find that quark flavor mixing in this theory is naturally hierarchical too, 
once one requires that the dynamically-generated quark masses are themselves hierarchical -- the light quark masses need not be tuned in this theory, being generated instead by the flavor-blind flavor-locking 
portal to the hierarchon sector -- and the mixing is generically of the observed size. The collider phenomenology of this theory is quite rich if the additional Higgs bosons are light, with testable signatures at the LHC or HL-LHC.

For an example benchmark point in the theory parameter space, we showed that this 
`flavor-locked flavorful 2HDM' model does not require significant tunings in order to reproduce the observed mass, CKM and meson mixing data.
In particular, $\mathcal{O}(1)$ variations in model parameters do not substantially or rapidly vary the agreement with the order of the observed CKM matrix, or, in other words, the hierarchical quark mixing is 
stable over $\mathcal{O}(1)$ variations in the parameters of the theory. By contrast, the SM features naively seven tunings: the five lighter quark masses, and the mixing angles $\theta_{23}$ and $\theta_{13}$ in 
the standard CKM parametrization, that produce small $|V_{cb}|$ and $|V_{ub}|$, respectively. 

The reduced amount of tuning of the quark masses and CKM mixing in the flavor-locked flavorful 2HDM 
does not come at the price of large NP contributions to meson mixing, even if the additional neutral Higgs bosons are light: $\mathcal{O}(1)$ variation of the flavor parameters does not lead to a significant deviation in meson mixing observables for heavy Higgs boson masses at around the electro-weak scale (e.g. $m_A \sim m_H \sim 500~\text{GeV})$ and moderate $\tan\beta$ (e.g. $\tan\beta \sim 5$), and may in fact better accommodate current meson mixing data than the SM itself. Further exploration of the flavor phenomenology of this theory
is left for future studies.

It is straightforward to extend this framework to the charged lepton sector. Possible ways to reproduce a realistic normal or inverted neutrino spectrum and the large neutrino mixing angles will be discussed elsewhere.

\section*{Acknowledgements}
We thank Simon Knapen for helpful conversations and for comments on the manuscript. WA and SG thank the Mainz Institute for Theoretical Physics (MITP) for its hospitality and support during parts of this work. The work of WA, SG and DR was in part performed at the Aspen Center for Physics, which is supported by National Science Foundation grant PHY-1607611.
The research of WA is supported by the National Science Foundation under Grant No. PHY-1720252.
SG is supported by a National Science Foundation CAREER Grant No. PHY-1654502.
We acknowledge financial support by the University of Cincinnati.

\appendix
\section{Analysis of the general flavon potential}
\label{app:GFPGM}
In this appendix we determine the global minimum of the flavon potential~\eqref{eqn:FP}.

\subsection{General flavon potential}
The single and pairwise field potentials~(\ref{eq:SP}), (\ref{eqn:SPFP}) are manifestly positive semidefinite. Noting that the $\mu_6$ terms can be written in the form $\Tr([\lmb^\pda_\al\lmb^\da_\be ]^\da\lmb^\pda_\al\lmb^\da_\be)$ and 
$\Tr([\lmb^\da_\al \lmb^\pda_\be]^\da\lmb^\da_\al \lmb^\pda_\be)$ and moreover that $\Tr [A^\da A] = \sum_{ij} |A_{ij}|^2 = 0$ if and only if $A =0$, the global minimum -- zero -- of $V_{\text{1f}}, V_{\text{2f}}$ is attained if and only if 
\begin{equation}
	\label{cond:AC}
\parbox{0.8\linewidth}{
\begin{enumerate}
	\item $\Tr[\lmbE{\pda}{\al} \lmbE{\da}{\al}] = r^2$,
	\item $ \lmbE{}{\al}  $ is rank-1,
	\item $\lmbE{\da}{\al} \lmbE{}{\be} = 0$ and $\lmbE{}{\al} \lmbE{\da}{\be} = 0$ for all $\al \not= \be$.
\end{enumerate}
}
\end{equation}
These algebraic conditions are equivalent to the set $\langle \lmb_\al \rangle$ being simultaneously real diagonalizable with disjoint unit rank spectra. That is, 
\begin{equation}
	\label{eqn:UTV}
	\lmbE{}{1}   = U\, \text{diag}\{r,0,0,\ldots\}\, V^\da\,, \qquad \lmbE{}{2}  = U\, \text{diag}\{0,r,0,\ldots\}\, V^\da\,, \ldots\,,
\end{equation}
with $U$, $V$ generic unitary matrices, the same for all $\lmb_\al$, that are flat directions of the global minimum, and $r$ real. A similar analysis follows immediately for the down-type potentials, so that 
\begin{equation}
	\label{eqn:DTV}
	\lmbE{}{\hat{1}}  = \hat{U}\, \text{diag}\{\hat{r},0,0,\ldots\}\, \hat{V}^\da\,, \qquad \lmbE{}{\hat{2}}   = \hat{U}\, \text{diag}\{0,\hat{r},0,\ldots\}\, \hat{V}^\da\,,\ldots\,.
\end{equation}
We refer to this type of aligned structure as `flavor-locked'. (It is possible to switch the rank-1 structure for degeneracy by setting $\mu_2<0$~\cite{Knapen:2015hia}, though we do not consider this possibility in this work.)

\subsection{Mixing terms: single flavon generation}
The first, $\nu_1$, term of the mixed potential~\eqref{eqn:VM} manifestly respects the vacuum of $V_{\text{1f}}$ and $V_{\text{2f}}$. It follows from the Cauchy-Schwarz inequality and positive semidefiniteness of $\lmb^\pda_\al\lmb^\da_\al$, that 
\begin{equation}
\Tr\big(\lmb^\pda_\al\lmb^\da_\al\big)\Tr\big(\lmb^\pda_\hal\lmb^\da_\hal\big) \ge \Tr\big(\lmb^\pda_\al\lmb^\da_\al\lmb^\pda_\hal\lmb^\da_\hal\big)\,.
\end{equation}
Hence for the case of $n=1$ generations of flavons, the $\nu_2$ term and full potential is immediately positive semidefinite, with global minimum at $V_{\text{fl}} = 0$.  Based on the flavor-locked configurations in Eqs.~\eqref{eqn:UTV} and~\eqref{eqn:DTV}, 
\begin{equation}
	[\langle\lmb^\da_\al \rangle\langle \lmb^\pda_\hal \rangle]_{I\hat J} = \mathcal{V}_{\text{ckm}}^{\al \hal}r \hat{r}\,\, [V]^\pda_{IJ} \delta_{\al J} \delta_{\hal \hat I} [\hat{V}]^\da_{\hat I \hat J}\,,
\end{equation}
in which we have momentarily restored the $U(N)_U\times U(N)_D$ indices and $\mathcal{V}_{\text{ckm}}= U^\da \hat{U}$ is the unitary CKM matrix. Without loss of generality, we can choose the non-zero eigenvalues of the single up and down flavon being in the first diagonal entry, at the flavor-locked configuration. One then obtains for the $n=1$ mixed potential
\begin{equation}
	V_{\text{mix}} = -\nu_2 r^2 \hat{r}^2 \Big[ \big|\mathcal{V}_{\text{ckm}}^{1\hat1}\big|^2  - 1\Big]\,.
\end{equation}
This vanishes if and only if $\mathcal{V}_{\text{ckm}}$ is $1 \oplus (N-1)$ block unitary, i.e.
\begin{equation}
	\label{eqn:SGCKM}
	\mathcal{V}_{\text{ckm}} = \begin{pmatrix} 1 & 0\\0 & \mathcal{V}_{N-1} \end{pmatrix}\,,
\end{equation}
in which $\mathcal{V}_{N-1}$ is an $N-1\times N-1$ unitary submatrix (as in Eq.~\eqref{eqn:MGCKMA}). Therefore, the potential has a global minimum if and only if the flavons lie in the flavor-locked configuration, with a block-unitary mixing matrix.

\subsection{Mixing terms: arbitrary flavon generations}
\label{app:AFG}
For the general case that $N \ge n\ge 1$, the $\nu_2$ term is not positive definite by itself. The full potential may, however, be reorganized into the form
\begin{equation}
	\label{eqn:FPA}
	V_{\text{fl}} =  \sum_\al U^{\al}_{\text{1f}} + \sum_{\al < \be} U^{\al\be}_{\text{2f}} + \sum_\hal U^{\hal}_{\text{1f}}  + \sum_{\hal < \hbe}U^{\hal\hbe}_{\text{2f}}  + U^0_{\text{mix}} + \sum_{\al,\,\hal}U^{\al\hal}_{\text{mix}}\,.
\end{equation}
in which the pure up-type potentials 
\begin{align}
U_{\text{1f}}^\al  
	& = \mu_1\Big| \Tr\big(\lmb_\al^\da \lmb_\al^\pda\big) - r^2 \Big|^2 + \bigg(\mu_2 + \frac{\nu_2}{2} \frac{\hat r^2}{r^2}\bigg)\bigg[ \big|\Tr\big(\lmb_\al^\da \lmb_\al^\pda\big)\big|^2 - \Tr\big(\lmb^\pda_\al \lmb^\da_\al \lmb^\pda_\al \lmb^\da_\al\big) \bigg]\,,\nn \\
U_{\text{2f}}^{\al\be} & = \mu_3\Big| \Tr\big(\lmb_\al^\da \lmb_\al^\pda\big) - \Tr\big(\lmb_\be^\da \lmb_\be^\pda\big)\Big|^2 + \mu_4 \Big|\Tr\big(\lmb_\al^\da \lmb_\be^\pda\big) \Big|^2\nn \\  
 	& \qquad + \mu_{6,1}\Tr\big(\lmb^\da_\al \lmb^\pda_\al\lmb^\da_\be \lmb^\pda_\be\big) + \bigg(\mu_{6,2} - \frac{\nu_2\hat r^2}{r^2}\bigg)\Tr\big(\lmb^\pda_\al \lmb^\da_\al\lmb^\pda_\be \lmb^\da_\be\big)\,,
\end{align}
and similarly for the down-type potentials, exchanging all unhatted and hatted couplings. The two mixed potentials
 \begin{align}
U^0_{\text{mix}} 
	& = \frac{\nu_2 r^2 \hat r^2}{2} \Tr \bigg[\bigg( \sum_\al \frac{\lmb^\pda_\al \lmb_\al^\da}{r^2}  - \sum_\hal \frac{\lmb^\pda_\hal \lmb_\hal^\da}{\hat r^2}\bigg)^2\bigg]\,,\label{eqn:U0} \\
U^{\al\hal}_{\text{mix}}	
	& = \bigg(\nu_1 - \frac{\nu_2}{2 n} \bigg)r^2 \hat r^2 \bigg| \Tr\big(\lmb_\al^\pda \lmb_\al^\da\big)/r^2 - \Tr\big(\lmb_\hal^\pda \lmb_\hal^\da\big)/\hat r^2\bigg|^2\,.
\end{align}
Hence each term of the full potential is now positive semidefinite, provided
\begin{equation}
	\label{eqn:CGM}
	\mu_{6,2} \ge \nu_2 \hat r^2/ r^2\,, \qquad \hat\mu_{6,2} \ge \nu_2 r^2/\hat r^2\,, \qquad \text{and} \qquad \nu_1 \ge \nu_2/(2 n)\,.
\end{equation}
We write the flavor-locked configuration in the ordered form of Eqs.~\eqref{eqn:UTV} and \eqref{eqn:DTV}, so that the first $n$ eigenvalues of $\lmbE{}{\al}$ are non-zero. At the flavor-locked configuration, the mixed potential becomes
\begin{equation}
	\sum_{\al,\hal}V^{\al\hal}_{\text{mix}} = -\nu_2 r^2 \hat{r}^2 \sum_{\al,\hal}\Big[ \big|\mathcal{V}_{\text{ckm}}^{\al \hal}\big|^2  - 1/n\Big] = 0\,.
\end{equation}
Unitarity ensures that 
\begin{equation}
 	\sum_{\al,\hal =1}^{n}\big|\mathcal{V}_{\text{ckm}}^{\al \hal}\big|^2 \le n\,,
\end{equation}
so that on the flavor-locked contour the mixing terms and hence full potential is minimized, with $V_{\text{fl}} = 0$, if and only if 
$\mathcal{V}_{\text{ckm}}$ is $n \oplus (N-n)$ block unitary. I.e.
\begin{equation}
	\label{eqn:MGCKM}
	\mathcal{V}_{\text{ckm}} = U^\da\hat{U} = \begin{pmatrix} \mathcal{V}_n & 0 \\ 0 &  \mathcal{V}_{N-n} \end{pmatrix}\,,
\end{equation}
with $\mathcal{V}_k$ a $k\times k$ unitary matrix. Note that the $n$ or $N-n$ block CKM rotations are flat directions of the global minimum, and therefore $\mathcal{V}_n$ and $\mathcal{V}_{N-n}$ may be any arbitrary unitary submatrices with generically $\mathcal O(1)$ entries. We often refer to Eq.~\eqref{eqn:MGCKM} in combination with Eqs.~\eqref{eqn:UTV} and~\eqref{eqn:DTV} as the `flavor-locked' configuration, too.

\subsection{Local minimum analysis}

So far we have shown that under the conditions~(\ref{eqn:CGM}) the global minimum of the potential is $V_\text{fl} = 0$ and it is realized if and only if the flavons are in the flavor-locked configuration. 
One may also explore the weaker condition that the flavor-locked configuration is only a local minimum of the potential, by applying the general perturbations 
\begin{equation}
	\lmbE{}{\al} \to \lmbE{}{\al} + \epsilon X_\al\,, \qquad \text{and} \qquad \lmbE{}{\hal} \to \lmbE{}{\hal} + \epsilon X_\hal\,.
\end{equation}
To this end, it is convenient to define
\begin{equation}
	H_\al = \frac{1}{r^2}\Big[\lmbE{\pda}{\al} X_\al^\da + X_\al^\pda \lmbE{\da}{\al} \Big]\,,\qquad P = \frac{1}{r^2} \sum_\al \lmbE{\pda}{\al}\lmbE{\da}{\al}\,,\qquad \hat P = \frac{1}{\hat r^2} \sum_\hal \lmbE{\pda}{\hal}\lmbE{\da}{\hal}\,,
\end{equation}
Observe $H_\al$ is Hermitian and $\Tr[P] = n$. One may show that $\Tr[P H_\al] = \Tr[H_\al]$, and, as a consequence of the block unitarity~\eqref{eqn:MGCKM}, that further $\Tr[\hat P H_\al] = \Tr[H_\al]$.  
Under perturbation of the mixing terms, one finds to $\mathcal{O}(\epsilon^2)$,  
\begin{multline}
	\delta [U^0_{\text{mix}} + \sum_{\al, \hal}U^{\al\hal}_{\text{mix}}]  = \epsilon^2\frac{\nu_2 r^2 \hat r^2}{2} \Tr\bigg[ \Big(\sum_\al H_\al - \sum_\hal H_\hal\Big)^2 \bigg] \\
			+ \epsilon^2\bigg(\nu_1 - \frac{\nu_2}{2 n}\bigg)r^2\hat r^2 \sum_{\al,\hal}\Big| \Tr H_\al -  \Tr H_\hal \Big|^2\,,
\end{multline}
which is positive semidefinite, provided the condition 
\begin{equation}
	\nu_1\ge \nu_2/(2n)\,,
\end{equation}
holds (cf. (\ref{eqn:CGM})). The vacuum configuration in (\ref{eqn:FLC}) is then a local minimum of the flavon potential.

More generically, one may also re-organize the potential, such that
\begin{equation}
	V_{\text{fl}} =  \bar U^0_{\text{1f}}  + \sum_\al \bar U^{\al}_{\text{1f}} + \sum_{\al < \be} \bar U^{\al\be}_{\text{2f}} + \sum_\hal \bar U^{\hal}_{\text{1f}}  + \sum_{\hal < \hbe}\bar U^{\hal\hbe}_{\text{2f}}  + \bar U^0_{\text{mix}} + \sum_{\al,\,\hal}\bar U^{\al\hal}_{\text{mix}}\,.
\end{equation}
in which we have defined, for an arbitrary real coefficient, $\omega$,
\begin{align}
\bar U^0_{\text{1f}}
	& = \omega\frac{\nu_2}{2n} \frac{\hat r^2}{r^2}\bigg|\sum_\al\big[\Tr\big(\lmb_\al^\da \lmb_\al^\pda\big) - r^2\big] \bigg|^2\nn\\
\bar U_{\text{1f}}^\al  
	& = \bigg(\mu_1 - \omega\frac{\nu_2}{2} \frac{\hat r^2}{r^2} \bigg)\Big| \Tr\big(\lmb_\al^\da \lmb_\al^\pda\big) - r^2 \Big|^2 + \bigg(\mu_2 + \frac{\nu_2}{2} \frac{\hat r^2}{r^2}\bigg)\bigg[ \big|\Tr\big(\lmb_\al^\da \lmb_\al^\pda\big)\big|^2 - \Tr\big(\lmb^\pda_\al \lmb^\da_\al \lmb^\pda_\al \lmb^\da_\al\big) \bigg]\,,\nn \\
\bar U_{\text{2f}}^{\al\be} & = \bigg(\mu_3 - (1-\omega)\frac{\nu_2}{2n} \frac{\hat r^2}{r^2} \bigg)\Big| \Tr\big(\lmb_\al^\da \lmb_\al^\pda\big) - \Tr\big(\lmb_\be^\da \lmb_\be^\pda\big)\Big|^2 + \mu_4 \Big|\Tr\big(\lmb_\al^\da \lmb_\be^\pda\big) \Big|^2\nn \\  
 	& \qquad + \mu_{6,1}\Tr\big(\lmb^\da_\al \lmb^\pda_\al\lmb^\da_\be \lmb^\pda_\be\big) + \bigg(\mu_{6,2} - \frac{\nu_2 \hat r^2}{r^2}\bigg)\Tr\big(\lmb^\pda_\al \lmb^\da_\al\lmb^\pda_\be \lmb^\da_\be\big)\,,
\end{align}
and analogously in the down sector for the $\hat\alpha$ and $\hat\beta$ pieces. The mixing terms are given by
 \begin{align}
\bar U^0_{\text{mix}} 
	& = \frac{\nu_2 r^2 \hat r^2}{2}\bigg\{ \Tr \bigg[\bigg( \sum_\al \frac{\lmb^\pda_\al \lmb_\al^\da}{r^2}  - \sum_\hal \frac{\lmb^\pda_\hal \lmb_\hal^\da}{\hat r^2}\bigg)^2 \bigg]  - \frac{1}{n} \bigg|\Tr\Big(\sum_\al \frac{\lmb^\pda_\al \lmb_\al^\da}{r^2}  - \sum_\hal \frac{\lmb^\pda_\hal \lmb_\hal^\da}{\hat r^2}\Big)\bigg|^2\bigg\},\nn \\
\bar U^{\al\hal}_{\text{mix}}	
	& =\nu_1 r^2\hat r^2  \big|\Tr\big(\lmb_\al^\pda \lmb_\al^\da\big)/r^2 - \Tr\big(\lmb_\hal^\pda \lmb_\hal^\da\big)/\hat r^2\big|^2\,.
\end{align}
This time, under perturbations of the flavor-locked configuration, one finds
\begin{equation}
	\delta \bar U^0_{\text{mix}} = \epsilon^2\frac{\nu_2 r^2 \hat r^2}{2}\Tr \bigg[\bigg( \sum_\al H_\al  - \sum_\hal H_\hal - \frac{P}{n} \Tr\Big[ \sum_\al H_\al  - \sum_\hal H_\hal \Big]\bigg)^2 \bigg]\,,
\end{equation}
which is positive semidefinite. Hence, no matter the form of the $\nu_1$ term, a local minimum can also be achieved for the case that
\begin{equation}
	\mu_1 \ge \omega\frac{\nu_2}{2} \frac{\hat r^2}{r^2}\,,\qquad \mu_3 \ge (1-\omega)\frac{\nu_2}{2n} \frac{\hat r^2}{r^2}\,,\qquad \omega \ge 0\,, \qquad \mu_{6,2} \ge \frac{\nu_2 \hat r^2}{r^2}\,,
\end{equation}
and similarly for the hatted couplings.

\subsection{Two-Higgs alignment conditions}
\label{app:THGFP}
The Two-Higgs potential~\eqref{eqn:THFP} is equivalent to the general potential~\eqref{eqn:FP}, but with the $t$--$c$, $t$--$u$ and $b$--$d$, $b$--$s$ cross-terms effectively vanishing.
The vacuum for $V_{\text{1f}} + V_{\text{2f}}$ then has the structure
\begin{equation}
\parbox{0.8\linewidth}{
\begin{enumerate}
	\item $\Tr[\langle \lmb_\al^\da \rangle \langle \lmb_\al^\pda\rangle] = r^2$,
	\item $\langle \lmb_\al \rangle $ is rank-1,
	\item $\langle \lmb_c^\da \rangle \langle \lmb^\pda_u \rangle = 0$ and $\langle \lmb_c^\pda \rangle \langle \lmb_u^\da\rangle = 0$
\end{enumerate}
}
\end{equation}
but neither $\langle \lmb_t^\da \rangle \langle \lmb^\pda_{c,u} \rangle$ nor $\langle \lmb_t^\pda\rangle \langle \lmb^\da_{c,u} \rangle$ need to vanish, and similarly for the down-type flavons. The potentials $V_{\text{fl}, \text{h}}$ and $V_{\text{fl},\text{l}}$ then each have a $N=3$ flavor-locked vacuum, with generation number $n=1$ and $n=2$, respectively. This leads immediately to the vacuum in eqs~\eqref{eqn:21FV} and~\eqref{eqn:UUH}.

\section{Flavor basis for the F2HDM Yukawa texture} \label{app:textures}
Starting from the general parametrization of the flavor-locked Yukawas in~(\ref{eqn:PPMM}) we perform the following quark field rotations in flavor space
\begin{equation}
 U_L \to \mathcal{U}_{U_L} U_L ~,~~ D_L \to \mathcal{U}_{D_L} D_L  ~,~~ U_R \to \mathcal{U}_{U_R} U_R ~,~~ D_R \to \mathcal{U}_{D_R} D_R ~, 
\end{equation}
where the $\mathcal{U}_i$ are $2\oplus 1$ block unitary matrices
\renewcommand*{\arraystretch}{.8}
\begin{eqnarray}
 && \mathcal{U}_{U_L} = 
 \begin{pmatrix} 
 \cos\theta_{U_L} & \sin\theta_{U_L} & 0 \\
 -\sin\theta_{U_L} & \cos\theta_{U_L} & 0 \\
 0 & 0 & 1
 \end{pmatrix} ~,~~
 \mathcal{U}_{D_L} = 
 \begin{pmatrix} 
 \cos\theta_{D_L} e^{i \psi_{D_L}} & \sin\theta_{D_L} & 0 \\
 -\sin\theta_{D_L} e^{i \psi_{D_L}} & \cos\theta_{D_L} & 0 \\
 0 & 0 & 1
 \end{pmatrix}~,\nn \\
 && \mathcal{U}_{U_R} = 
 \begin{pmatrix} 
 \cos\theta_{U_R} & \sin\theta_{U_R} & 0 \\
 -\sin\theta_{U_R} & \cos\theta_{U_R} & 0 \\
 0 & 0 & 1
 \end{pmatrix} ~,~~
 \mathcal{U}_{D_R} = 
 \begin{pmatrix} 
 \cos\theta_{D_R} & \sin\theta_{D_R} & 0 \\
 -\sin\theta_{D_R} & \cos\theta_{D_R} & 0 \\
 0 & 0 & 1
 \end{pmatrix} ~.
\end{eqnarray}
The rotation angels and the phase are chosen such that
\begin{eqnarray}
 \tan\theta_{U_L} &=& \sin\theta_{13} \tan\theta_{23} ~, \\
 \tan\theta_{U_R} &=& \sin\vartheta_{13} \tan\vartheta_{23} ~, \\
 \tan\theta_{D_R} &=& \sin\hat\vartheta_{13} \tan\hat\vartheta_{23} ~, \\
 \tan\theta_{D_L} &=& \sin\theta_{13} \tan\theta_{23} \cos\psi_{D_L} - \tan\theta \frac{\cos\theta_{13}}{\cos\theta_{23}} \cos(\psi_m + \psi_{D_L}) ~, \\
 \tan\psi_{D_L} &=& \frac{\tan\theta \sin\psi_m}{\sin\theta_{23} \tan\theta_{13} - \tan\theta \cos\psi_m}  ~.
\end{eqnarray}
In this flavor basis the Yukawas in~(\ref{eqn:PPMM}) reproduce the F2HDM textures from Eq.~(\ref{eqn:Ys}) with coefficients that depend on the several angles $\theta_{13},\theta_{23},\vartheta_{13},\vartheta_{23},\hat\vartheta_{13},\hat\vartheta_{23},\theta$ and phases $\psi_d,\psi_s,\psi_u,\psi_c,\psi_m$. 

\bibliography{FLF2hDM}

\end{document}